\documentclass[aps,prb,twocolumn,superscriptaddress,longbibliography]{revtex4-2}
\usepackage{amsmath,amssymb}
\usepackage[pdftex]{hyperref,graphicx}
\hypersetup{colorlinks = true, urlcolor = blue, linkcolor = blue, citecolor = blue}
\usepackage{physics}
\usepackage{xcolor}
\usepackage{bm}
\usepackage[section]{placeins} 

\begin{document}
\title{Density of states, transport, and topology in disordered Majorana nanowires}
\author{Sankar Das Sarma}
\affiliation{Condensed Matter Theory Center and Joint Quantum Institute, Department of Physics, University of Maryland, College Park, Maryland 20742, USA}
\author{Haining Pan}
\affiliation{Condensed Matter Theory Center and Joint Quantum Institute, Department of Physics, University of Maryland, College Park, Maryland 20742, USA}

\begin{abstract}
    Motivated by a recent breakthrough transport experiment {[Phys. Rev. B.107.245423 (2023)]} in Majorana nanowires, we theoretically investigate local and nonlocal transport in Majorana nanowires in various disorder regimes, correlating the transport properties with the corresponding local and total density of states as well as various topological diagnostics.  We find three distinct disorder regimes, with weak (strong) disorder regimes manifesting (not manifesting) topological superconductivity with clear end Majorana zero modes for longer (but not necessarily for shorter) wires. The intermediate disorder regime is both interesting and challenging because the topology depends on many details in addition to the strength of disorder, such as the precise disorder configuration and the wire length.  The intermediate disorder regime often manifests multiple effective transitions between topological and nontopological phases as a function of system parameters (e.g., the Zeeman field), and is consistent with the recent Microsoft experiment reflecting small topological gaps and narrow topological regimes in the parameter space.
\end{abstract}

\maketitle

\section{Introduction}\label{sec:introduction}

A recent Microsoft experiment~\cite{microsoftquantum2023inasal} represents a breakthrough in the study of topological superconductivity in Majorana nanowires composed of semiconductor-superconductor (SM-SC) hybrid structures.  In particular, the experiment reports detailed measurements of both local and nonlocal tunneling conductances, i.e., the full four-component conductance matrix,  in many nanowire samples as functions of bias voltage, gate voltage, cutter voltage, and magnetic field, thus producing detailed operational phase diagrams for the existence (or not) of topological superconductivity with a finite topological gap as a function of sample, magnetic field, and gate voltage. The extracted topological gap from the measured nonlocal conductance is typically small ($\sim$ 25 $\mu$eV), much less than the expected pristine proximity-induced SC gap, and not all samples manifest topological superconductivity.  The topological regime with a finite topological gap, where it exists, is typically rather small in the magnetic-field and gate-voltage variations, indicating limited robustness of the topological superconductivity, most likely arising from the unavoidable presence of unintentional random disorder in the system~\cite{ahn2021estimating,woods2021chargeimpurity,dassarma2021disorderinduced,pan2020physical,liu2012zerobias,bagrets2012class,akhmerov2011quantized,sau2013density,brouwer2011probability}. The hallmark of this experiment (which has never been achieved before experimentally) is the simultaneous observation of correlated zero-bias conductance peaks (ZBCP) from both wire ends in the local tunneling spectroscopy and the manifestation of approximate bulk gap closing/opening in the nonlocal tunneling spectroscopy, as is expected for Majorana-carrying topological systems~\cite{pan2021threeterminal,lai2019presence,rosdahl2018andreev}. The experiment extracts a topological gap from a careful analysis of the conductance spectroscopic measurements of all four components of the tunnel conductance as functions of magnetic field and gate voltages.

Given the great importance of this breakthrough experiment~\cite{microsoftquantum2023inasal}, which supersedes all earlier Majorana nanowire experiments and the first Majorana measurement reporting topological superconductivity~\cite{dassarma2023search}, we provide in this paper the minimal theoretically expected results for the local and nonlocal conductance in Majorana nanowires as a function of disorder and wire length along with a number of relevant theoretical quantities in order to better understand the role of disorder in the experiment as well as to put the experiment in the proper context. In particular, we calculate the total and local density of states as well as several theoretical diagnostics for topology and Majorana zero modes such as the transport topological invariant~\cite{akhmerov2011quantized,fulga2011scattering, fulga2012scattering,dassarma2016how}, the thermal conductance~\cite{akhmerov2011quantized,fulga2011scattering}, and the Majorana localization length (which is effectively the SC coherence length)~\cite{cheng2009splitting}. We emphasize that experimentally only the tunnel conductance matrix is measured directly, and our calculations of the other theoretical quantities enable a deeper understanding of the underlying topology as compared with the conductance. A comparison among all these results, presented in Sec.~\ref{sec:model} of this paper, gives a nuanced view of the complicated interplay among disorder, topology, finite wire length, and superconductivity, showing that finite nanowires in the intermediate disorder regime (which is likely to be the regime of current experimental interest) is a rather complicated system to understand using minimal physical pictures.  It appears that longer wires and less disorder are essential in obtaining a clear decisive interpretation of the experimental results, although we do find that the intermediate-disorder regime allows the existence of fragile topological regimes with small gaps existing over small regimes of experimental parameters as claimed in Ref.~\cite{microsoftquantum2023inasal}.  A related study has recently explored the effective topological phase diagram in depth in the intermediate-disorder regime using realistic sample disorder~\cite{sarma2023spectral}, and our paper is complementary to this study. A significant difference between our study and this other study is that we use short-range random disorder characterized only by the disorder strength, whereas the other study uses more realistic long-range disorder, which needs two parameters (disorder strength and correlation length) to be characterized.

The rest of this paper is organized as follows.  In Sec.~\ref{sec:model}, we provide the basic theory, describe our calculations, and present our detailed numerical results.  We conclude in Sec.~\ref{sec:discussion} with a detailed discussion of our results in the context of the Microsoft experiment~\cite{microsoftquantum2023inasal}.

\section{Model, theory, and results}\label{sec:model}
We use the minimal model for the SM-SC hybrid nanowire system, where a one-dimensional (1D) SM nanowire is proximitized by a SC, leading to an induced SC pairing. The nanowire itself has Rashba spin-orbit (SO) coupling and Zeeman spin-splitting terms in its Hamiltonian.  So, the pristine system is described by the minimal Hamiltonian $H_\text{tot}$ describing the basic system originally introduced in Refs.~\onlinecite{sau2010robustness,sau2010generic,lutchyn2010majorana,oreg2010helical,dassarma2023search} 

\begin{equation}\label{eq:H_tot}
    H_\text{tot}=\left( -\frac{\hbar^2}{2m^*}\partial_x^2-i\alpha \partial_x\sigma_y-\mu \right)\tau_z+V_\text{Z}\sigma_x+\Delta\tau_x.
\end{equation}

In Eq.~\eqref{eq:H_tot}, each term is unambiguous, respectively representing the 1D kinetic energy, the SO coupling, the chemical potential (defined by the 1D carrier density and the applied gate voltage in the nanowire), the Zeeman splitting, and the induced gap.  The $\vec{\tau}$ and $\vec{\sigma}$ in Eq.~\eqref{eq:H_tot} are the $2\times2$ matrices representing the Nambu electron-hole SC pairing operator and the Pauli spin operator, respectively.  
Note that in principle, the various contributions (SO coupling $\alpha$, Zeeman splitting $V_{\text{Z}}$, chemical potential $\mu$, SC gap $\Delta$) to $H_\text{tot}$ are experimentally controlled, but for a given sample $\alpha$ and $\Delta$ are fixed whereas $\mu$ and $V_{\text{Z}}$ are controlled experimentally by the applied gate voltage and the applied magnetic field respectively. 
To further realistically simulate the coupling between the superconductor and semiconductor, we replace the constant superconducting term $\Delta\tau_x$ with a self-energy term $-\gamma \frac{\omega+\Delta_0\tau_x}{\sqrt{\Delta_0^2-\omega^2}}$, where $\gamma$ represents the effective coupling strength, $\Delta_0 $ is the gap of parent superconductor at zero field, and $\omega$ denotes the energy. However, the self-energy term only quantitatively suppresses the superconducting gap by reducing the effective superconducting gap at zero bias from $\Delta$ to $\gamma$, making our numerical result more similar to the experimental results, but does not qualitatively affect the formation of low-energy states. 
In addition to the Hamiltonian parameters entering Eq.~\eqref{eq:H_tot}, there are two other experimental parameters to consider: the temperature ($T$) and the wire length ($L$).  We take $T=0$ throughout since the experimental temperature is typically 20 mK ($\sim$ 2 meV), which is much lower than all the other energy scales in Eq.~\eqref{eq:H_tot}. We vary $L$ in order to discern the finite-length effects on Majorana nanowire physics.
We note that our theoretical length $L$ may not precisely be the experimental wire length, which is determined by the details of the experimental structures, including the locations and actions of the various gates in the sample. The applicable length $L$ of the experimental wire is likely to be shorter than the nominal physical length.
{Furthermore, although a more detailed modeling of the experimental devices is possible in principle~\cite{lutchyn2011search,stanescu2011majorana,pekerten2017disorderinduced}, such as including multi-orbital effects; it is not necessary in this paper, as we aim to provide a minimal model that can demonstrate the physics of different disorder regimes for short and long wires.
In particular, experiments try to be in the one-subband (i.e., one orbital) limit because this maximizes the induced gap, and in general, the experimental parameters are unknown, making such a multi-subband simulation not to be particularly useful since it introduces additional unknown parameters in the theory.
Therefore, it suffices to choose 1D single-band model with only the essential ingredients included.}

The free fermion Majorana nanowire model, as described by Eq.~\eqref{eq:H_tot}, has been extensively studied theoretically~\cite{dassarma2023search} since it was first introduced in Refs.~\onlinecite{sau2010robustness,sau2010generic,lutchyn2010majorana,oreg2010helical}, and leads to a {topological quantum phase transition (TQPT)} for $V_\text{Z} =V_{\text{Z}c} = ( \Delta^2 + \mu^2)^{1/2}$  with $V_{\text{Z}} > V_{\text{Z}c} $ ($V_{\text{Z}} < V_{\text{Z}c} $) being the topological (trivial) SC phase, and the topological SC gap for $V_{\text{Z}}>V_{\text{Z}c} $ being proportional to the SO coupling strength $\alpha$. For long wires, where $L$ is much larger than the SC coherence length, the topological SC phase with $V_\text{Z}>V_{\text{Z}c}$ hosts non-Abelian MZMs at the wire ends. However, the situation is complex for short wires due to the ill-defined topology resulting from the strong hybridization between the two end MZMs~\cite{dassarma2012splitting}. Topology applies only when the wire length is larger than the SC coherence length (or, equivalently, the Majorana localization length), which is inaccessible experimentally (and is likely to be large for small induced SC gaps). Thus, although there is no question about the emergence of topological SC and the associated localized non-Abelian MZMs in long pristine nanowires in the SM-SC hybrid systems, the issue of what happens in realistic short wires ($L \sim 1$ micron) where most current experiments are conducted remains open, particularly since the SC coherence length in the experimental samples is unknown.  An additional complication arises from the fact that the bulk SC in the Al film inducing the proximity effect is quenched by Pauli blockade for a magnetic field $B \sim B_c\sim ~ 2-3$ T because Al spins align at this Clogston limit~\cite{clogston1962upper}, disallowing $s$-wave pairing.  If $V_{\text{Z}c}$ for TQPT is above or around this bulk SC Clogston limit, the $\Delta $ term in Eq.~\eqref{eq:H_tot} vanishes, and there is no Majorana physics in the problem.  In addition to the problems of finite (and perhaps small) $L$ and finite (and perhaps small) $ B_c$, which complicate experiments even in (hypothetical) pristine samples, there is the much bigger problem of disorder in real samples, which has hampered progress in the subject. In fact, we have argued elsewhere that all published Majorana nanowire experiments are dominated by disorder and not by intrinsic topological SC~\cite{pan2020physical,dassarma2021disorderinduced,dassarma2023search}.

The most detrimental problem, which is not included in Eq.~\eqref{eq:H_tot} representing pristine disorder-free systems, is the inevitable existence of disorder in real samples.  We include disorder in Eq.~\eqref{eq:H_tot} by modifying the pristine chemical potential to an effective chemical potential including a random potential $V(x)$, where $V(x)$ denotes a disorder potential added to the chemical potential.  From prior studies, it is now known that this chemical potential disorder is the most important disorder in experimental samples, and therefore we include only a random $V(x)$ in the chemical potential~\cite{pan2020physical, pan2021threeterminal,dassarma2021disorderinduced,pan2022ondemand,dassarma2023search}.

Since not much is known about the actual effective disorder in the SM-SC nanowire samples (except that disorder exists), we make the simplest approximation for $V(x)$ assuming it to be an uncorrelated random Gaussian distribution with zero mean and a variance of $\sigma^2$, where $\sigma$ then defines the effective disorder strength,

\begin{equation}\label{eq:V}
    V(x)=  \mathcal{N} (0, \sigma^2).
\end{equation}       

Note that $V(x)$ should be added to $\mu$ so that the corresponding term in Eq.~\eqref{eq:V} becomes $\mu - V(x)$.  The uncorrelated nature of our choice for disorder is somewhat artificial since a natural correlation length is imposed by the lattice cutoff (10 nm here) we introduce in diagonalizing Eq.~\eqref{eq:H_tot}. The typical lattice cutoff is of the order of $10-30$ nm, which is approximately consistent with the microscopic disorder estimates in SM-SC systems~\cite{ahn2021estimating,woods2021chargeimpurity}. One advantage of our model is that the problem depends on only one disorder parameter, strength, allowing a better characterization of disorder effects. One should think of Eq.~\eqref{eq:V} simply as a model of disorder, with values of the disorder strength $\sigma$ chosen to correspond approximately to the disorder in SM-SC nanowires, but we make no efforts to compare with the experimental data, instead focusing on how changing disorder from weak to strong modifies the theoretical results for a qualitative understanding.

The parameters chosen for our calculations corresponding approximately to experimental systems are as follows: the effective mass $m$ is $0.015~m_e$, the chemical potential $\mu$ is 1 meV, the SOC strength $\alpha$  is 0.5 eV\AA, the parent SC gap $\Delta_0$ is 0.2 meV, the effective coupling between SC and SM $\gamma$ is 0.2 meV, the disorder strength $\sigma$ is 0.3 meV for weak disorder, $2.5 - 5$ meV for intermediate disorder, and $10-30$  meV for strong disorder,  $T=0$, and $L$ varies from $0.5$ microns to 10 microns as shown in our results.  With these choices, the pristine TQPT is at $V_{\text{Z}c} = 1.41$ meV.  We assume one subband occupancy in the nanowire. 
We note that our choice for the disorder strength in our short-range model is consistent with the corresponding long-range Coulomb disorder choice in Ref.~\cite{sarma2023spectral} because our disorder of $\sim$ 3 meV for our discretization scale of 10 nm in Eq.~\eqref{eq:H_tot}  corresponds roughly to the correlated disorder of 0.5 meV used in Ref.~\cite{sarma2023spectral}.
We calculate the transport properties using Blonder-Tinkham-Klapwijk formalism~\cite{blonder1982transition}, which constructs the S matrix for the scattering process of an incoming electron at energy $E$ at each lead in the system. The numerics of calculation of the S matrix is performed with the help of a Python package Kwant~\cite{groth2014kwant}. To simulate the measurements in the experiment, i.e., the four components of the tunneling conductance matrix, we calculate the local differential conductance ($G_{\text{LL}}=\frac{\partial I_{\text{L}}}{\partial V_{\text{L}}}$ and $G_{\text{RR}}= \frac{\partial I_{\text{R}}}{\partial V_{\text{R}}}$) by tracing out the S matrix describing the process of Andreev reflection at the lead $i$, i.e., $G_{ii}=\frac{2e^2}{h} \tr\left( \left[ S_{ii}^{he} \right]^\dagger S_{ii}^{he} \right)$, and nonlocal differential conductance ($G_{\text{LR}}=-\frac{\partial I_{\text{L}}}{\partial V_{\text{R}}} $ and $G_{\text{RL}} = -\frac{\partial I_{\text{R}}}{\partial V_{\text{L}}}$) by tracing out the S matrices describing the process of the net transmission from lead $i$ to the other lead $j$, i.e., $G_{ij}=\frac{e^2}{h} \tr\left( \left[ S_{ij}^{ee} \right]^\dagger S_{ij}^{ee} - \left[ S_{ij}^{he} \right]^\dagger S_{ij}^{he} \right)$.  We calculate the conductance spectrum as a function of the bias voltage $V_{\text{bias}}$ and Zeeman field $V_{\text{Z}}$ for each disorder configuration without any ensemble averaging.
We make no efforts to compare with experiments; instead, we focus on broad trends in the results as functions of disorder ($\sigma$) and wire length ($L$) in order to better understand the role of crossover physics, induced by both disorder and wire length, in Majorana nanowires, for a qualitative interpretation of what may be going on in the experiment.  We do not explicitly put a cutoff $B_c$ (arising from the bulk gap collapse) in our theory because the Microsoft experiment is manifestly in the $B<B_c$ regime.  It should be assumed that the topological SC and Majorana physics simply disappear for $B>B_c$.

A salient feature of our paper is the calculation of several properties together, comparing them with each other.  In addition to the local and nonlocal conductance matrix, we also calculate two topological properties: the topological visibility $Q=\det(S_{\text{LL}})=\det(S_{\text{RR}})$~\cite{akhmerov2011quantized,fulga2011scattering, fulga2012scattering,dassarma2016how} and the thermal conductance $\kappa=\kappa_0\tr(S_{LR}S_{LR}^\dagger)$ ($i\neq j$)~\cite{akhmerov2011quantized,fulga2011scattering}, measured from both ends of the wire. 
These properties define the topology and the TQPT, respectively --- the system is in the topological (trivial) phase for $Q < 0$ ($Q > 0$)~\cite{akhmerov2011quantized,fulga2011scattering, fulga2012scattering,dassarma2016how}, and at the TQPT, the thermal conductance $\kappa$ is quantized to be $\kappa_0=\frac{\pi^2 k_B^2 T}{6h}$~\cite{akhmerov2011quantized,fulga2011scattering} (where $T $ is temperature, $k_B$ is the Boltzmann constant, and $h$ is the Planck constant). We note that in our theory, we calculate $\kappa/\kappa_0$ directly, which is well-defined in the $T=0$ limit. By comparing the calculated conductance (the only experimentally accessible property) with the topological diagnostics, we are able to provide key insights into the behavior of the finite disordered system.  

In addition to the two topological properties, we also calculate the Majorana localization length, which is equivalent to the SC coherence length ($\xi$), directly from the ground-state wavefunction at a specific $V_\text{Z}$, as shown in Appendix~\ref{app:localization}. To do this, we first obtain the wavefunction by diagonalizing the discretized version of the continuum Hamiltonian Eq.~\eqref{eq:H_tot} in the fermionic basis by replacing the differential operator $\partial_x \psi(x)$ with the finite difference $\frac{\psi(x+a)-\psi(x-a)}{2a}$, where the lattice constant $a$ is set to be 10 nm. We then decompose the ground-state wavefunction from the fermionic basis into two Majorana basis. We assume that the wavefunction in the Majorana basis, if localized at wire ends, has an exponential decay in the form of $\sim\exp(-x/\xi)$ for the left end or $\sim\exp((x-L)/\xi)$ for the right end (where $L$ is the wire length). Finally, we select the larger $\xi$ extracted from the wavefunctions in the Majorana basis from both ends.
The Majorana localization length (SC coherence length) serves to estimate $L/\xi$ to determine if the system is in the long wire regime ($L\gg\xi$), which is necessary for the topological behavior of the MZMs.

Finally, we calculate the total density of states (DOS), $\rho(E)$, as a function of the energy $E$ to identify subgap states in the wire. To more precisely determine the position of the localized subgap state, we also calculate the local density of states (LDOS), $\rho(E, x)$, as a function of both energy $E$ and position $x$. This allows us to directly visualize whether the subgap state is located at the ends or in the bulk of the wire, which is often the case in the presence of disorder due to the Griffiths effects~\cite{motrunich2001griffiths}. The results of the LDOS calculations can be found in Appendix~\ref{app:LDOS}.

Motivated by the Microsoft experiment, we consider different values of the wire length and disorder strength so that we can provide insights into the topological regimes of the experimental samples.  We consider $L= 0.5-10$ microns (from short to long limits) and disorder strengths of 0.3 meV (weak disorder), 2-3 meV (intermediate disorder), and 10-30 meV (strong disorder).  We deliberately vary the disorder strength by orders of magnitude so that the three regimes are qualitatively different, with most results focusing on the intermediate disorder, which we believe describes the Microsoft experiments. All parameters other than the wire length and the disorder strength are kept the same in all our results.

In the following sections of numerical results, we begin with a pristine no-disorder situation as shown in Figs.~\ref{fig:L_pris_1.0} and~\ref{fig:L_pris_3.0} to review what one should expect to observe for the ideal topological Majorana physics. 
Next, in Figs.~\ref{fig:L_muVar0.3_1.0} and~\ref{fig:L_muVar0.3_3.0}, we present the weak disorder situation to demonstrate the robustness of the MZM against weak disorder in Sec.~\ref{sec:weak}. Both pristine and weak disorder situations manifest topological superconductivity with MZMs.
Then, in Sec.~\ref{sec:intermediate}, we provide two sets of intermediate disorder configurations (Figs.~\ref{fig:L_muVar2.5_0.5}-\ref{fig:L_muVar2.5_10.0} and Figs.~\ref{fig:L_muVar3.0_1.0}-\ref{fig:L_muVar3.0_3.0}) for different wire lengths ranging from short wires ($L=0.5~\mu$m) to long wires ($L= 10~\mu$m) to show the complicated topology that depends on many details, such as wire length and disorder configurations.  
Finally, we provide examples in the presence of strong disorder in Sec.~\ref{sec:strong}, where almost no signals in the local conductance and nonlocal conductance spectra are observed, indicating the absence of any topological superconductivity.
Results presented in the Appendix complement the results in the main text, enhancing and clarifying the physics, and are referred to in the main text as appropriate.

\subsection{Pristine case}\label{sec:pristine}

\begin{figure*}[htbp]
    \centering
    \includegraphics[width=6.8in]{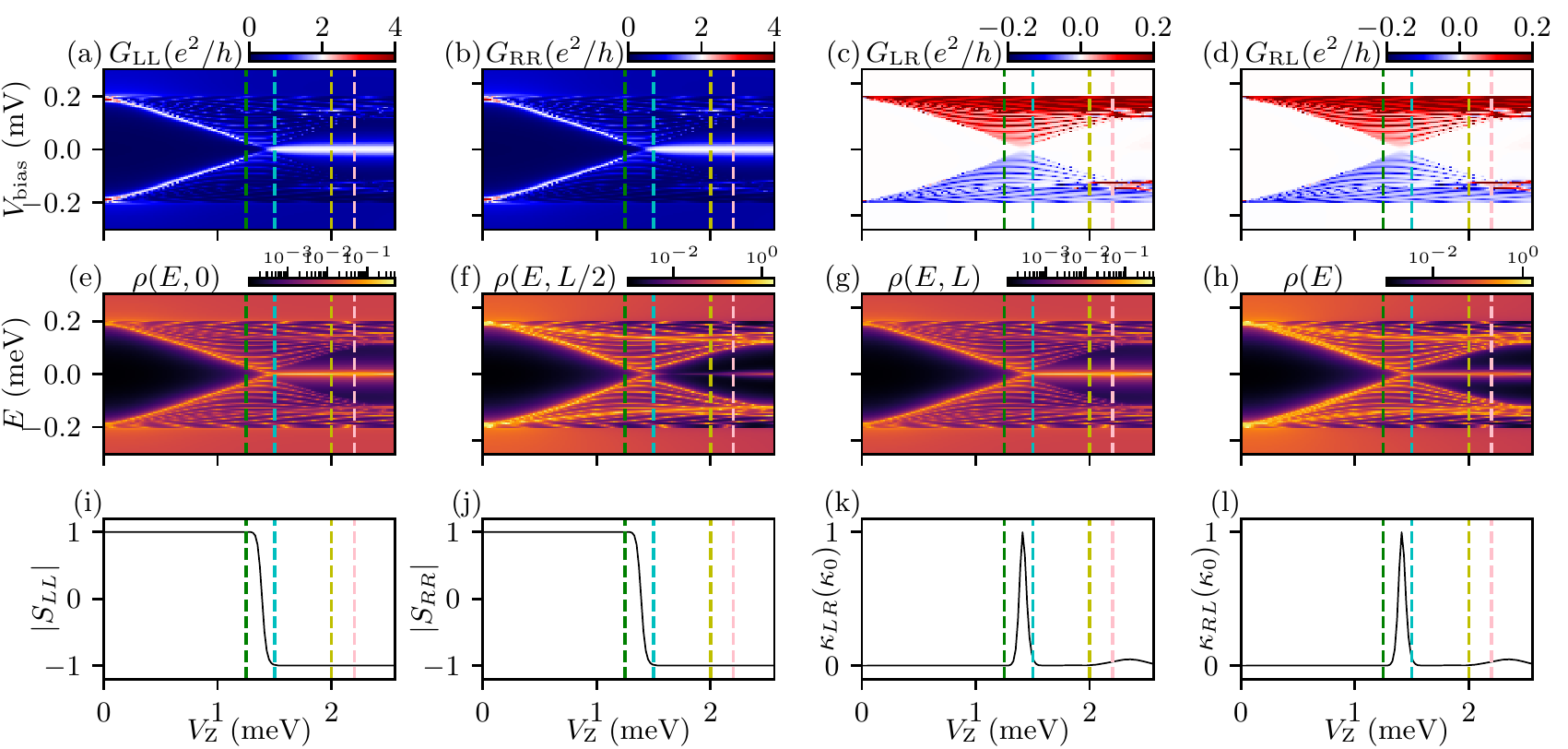}
    \caption{One-micron pristine wire. 
    (a)-(d) show the local and nonlocal conductances;
    (e)-(h) show the LDOS at $x=0, L/2, L$, and total DOS, respectively; Additional LDOS results in the bulk of the wire are presented in Fig.~\ref{fig:L_pris_1.0_LDOS} (Appendix~\ref{app:LDOS});
    (i)-(l) shows the topological visibility and thermal conductance from both ends. 
    The corresponding wavefunctions and their localization lengths at four typical $V_\text{Z}$ values, as indicated by the colored dashed line, are presented in Fig.~\ref{fig:wf_pris_1.0} (Appendix~\ref{app:localization}).}
    \label{fig:L_pris_1.0}
\end{figure*}

\begin{figure*}[htbp]
    \centering
    \includegraphics[width=6.8in]{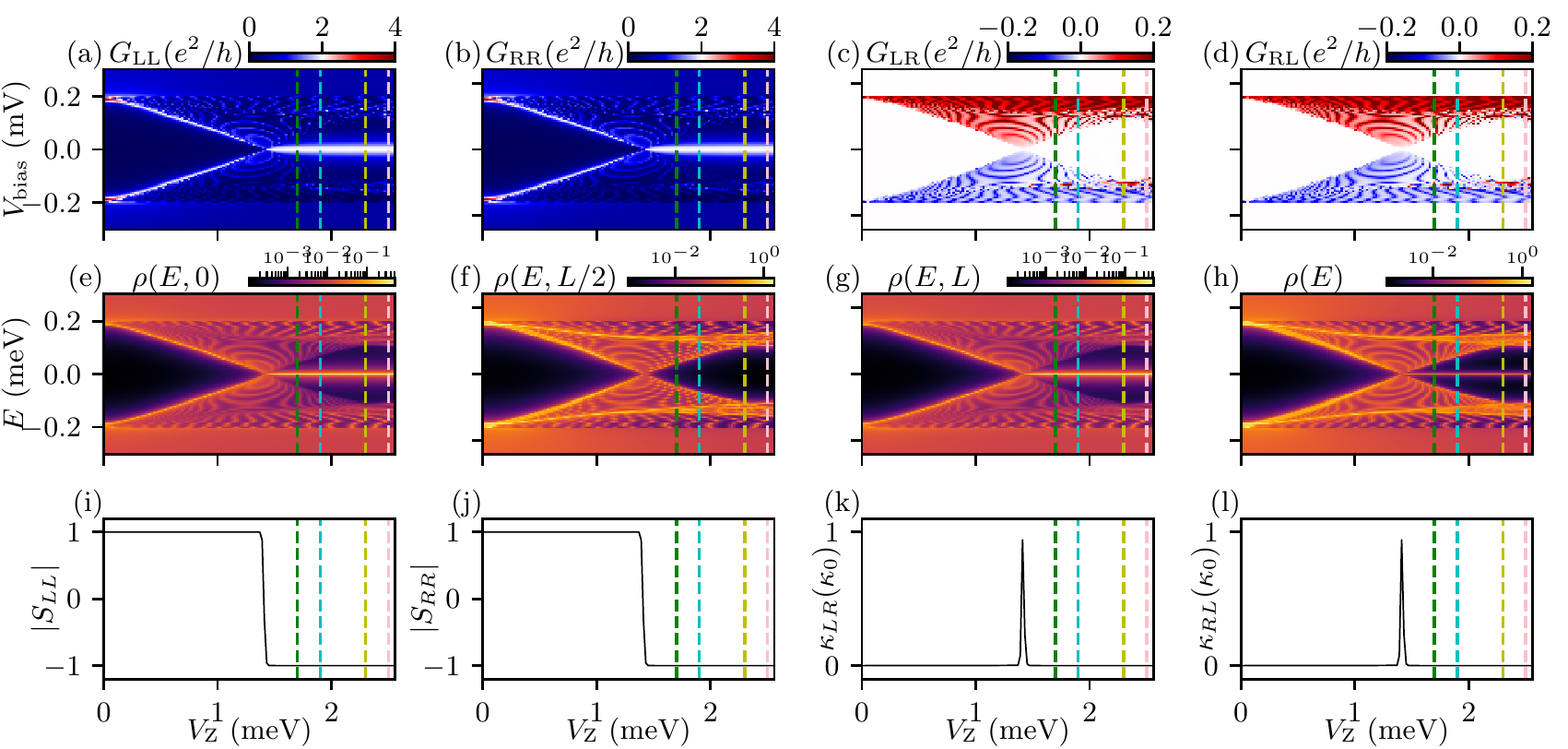}
    \caption{Three-micron pristine wire. 
    (a)-(d) show the local and nonlocal conductances;
    (e)-(h) show the LDOS at $x=0, L/2, L$, and total DOS, respectively; Additional LDOS results in the bulk of the wire are presented in Fig.~\ref{fig:L_pris_3.0_LDOS} (Appendix~\ref{app:LDOS});
    (i)-(l) shows the topological visibility and thermal conductance from both ends. 
    The corresponding wavefunctions and their localization lengths at four typical $V_\text{Z}$ values, as indicated by the colored dashed line, are presented in Fig.~\ref{fig:wf_pris_3.0} (Appendix~\ref{app:localization}).}
    \label{fig:L_pris_3.0}
\end{figure*}

Figures~\ref{fig:L_pris_1.0}(a) and~\ref{fig:L_pris_1.0}(b) show the calculated local conductance of a relatively short Majorana nanowire ($L = 1$  micron) in the pristine limit, where a ZBCP appears beyond the TQPT ($V_\text{Z}>1.41$ meV), indicating the presence of Majorana zero modes localized only at wire ends, as shown in Figs.~\ref{fig:L_pris_1.0}(e) -~\ref{fig:L_pris_1.0}(g). 
In Figs.~\ref{fig:L_pris_1.0}(c) and~\ref{fig:L_pris_1.0}(d), we observe an apparent gap closing and reopening below and above the TQPT at $V_\text{Z}=1.41$ meV in the nonlocal conductance spectra, serving as the basis for identifying the putative TQPT in recent experiments conducted by Microsoft~\cite{microsoftquantum2023inasal}. The topological visibility in Figs.~\ref{fig:L_pris_1.0}(i) and~\ref{fig:L_pris_1.0}(j) and thermal conductance in Figs.~\ref{fig:L_pris_1.0}(k) and~\ref{fig:L_pris_1.0}(l) also exhibit clear transitions from +1 to -1 and a peak quantized at $\kappa_0$, respectively, at the TQPT ($V_\text{Z}=1.41$ meV). All of these are typical features of Majorana zero modes in the pristine limit.

Figure~\ref{fig:L_pris_3.0} illustrates similar features observed in a longer pristine Majorana nanowire with $L=3$ microns, which exhibits a sharper transition due to the longer effective length $L/\xi$. The corresponding wavefunctions and their localization length $\xi$ at several specific $V_\text{Z}$ (indicated by dashed lines), presented in Figs.~\ref{fig:wf_pris_1.0} and~\ref{fig:wf_pris_3.0} (Appendix~\ref{app:localization}), respectively, further confirm the localization of Majorana zero modes at the wire ends and the absence of low-energy states in the bulk of the wire.

The purposes of presenting the pristine results are the following: (1) establishing a baseline in understanding the realistic results in the disorder samples; (2) establishing that even $L=1$ micron should be adequate for the manifestation of topological superconductivity and MZMs in the samples of interest in the disorder-free situation; and (3) establishing the salient topological features expected at the TQPT and in the topological phase.  The most important aspect of these pristine results is that the tunnel conductance and topological properties are consistent with each other as well as with the density of states.  

\subsection{Weak disorder}\label{sec:weak}

\begin{figure*}[htbp]
    \centering
    \includegraphics[width=6.8in]{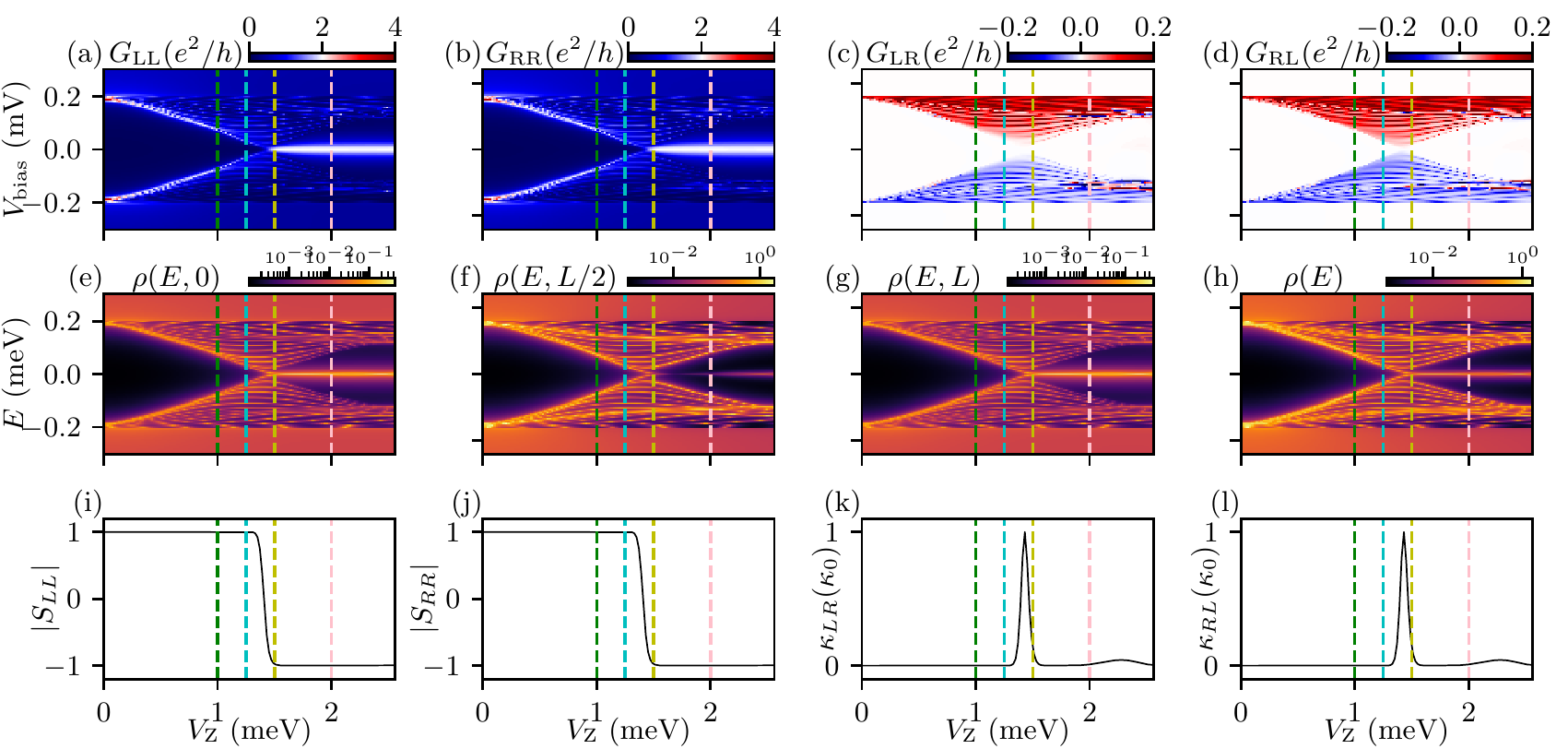}
    \caption{One-micron wire in the presence of weak disorder ($\sigma=0.3$ meV). 
    (a)-(d) show the local and nonlocal conductances;
    (e)-(h) show the LDOS at $x=0, L/2, L$, and total DOS, respectively; Additional LDOS results in the bulk of the wire are presented in Fig.~\ref{fig:L_muVar0.3_1.0_LDOS} (Appendix~\ref{app:LDOS});
    (i)-(l) shows the topological visibility and thermal conductance from both ends. 
    The corresponding wavefunctions and their localization lengths at four typical $V_\text{Z}$ values, as indicated by the colored dashed line, are presented in Fig.~\ref{fig:wf_muVar0.3_1.0} (Appendix~\ref{app:localization}). }
    \label{fig:L_muVar0.3_1.0}
\end{figure*}

\begin{figure*}[htbp]
    \centering
    \includegraphics[width=6.8in]{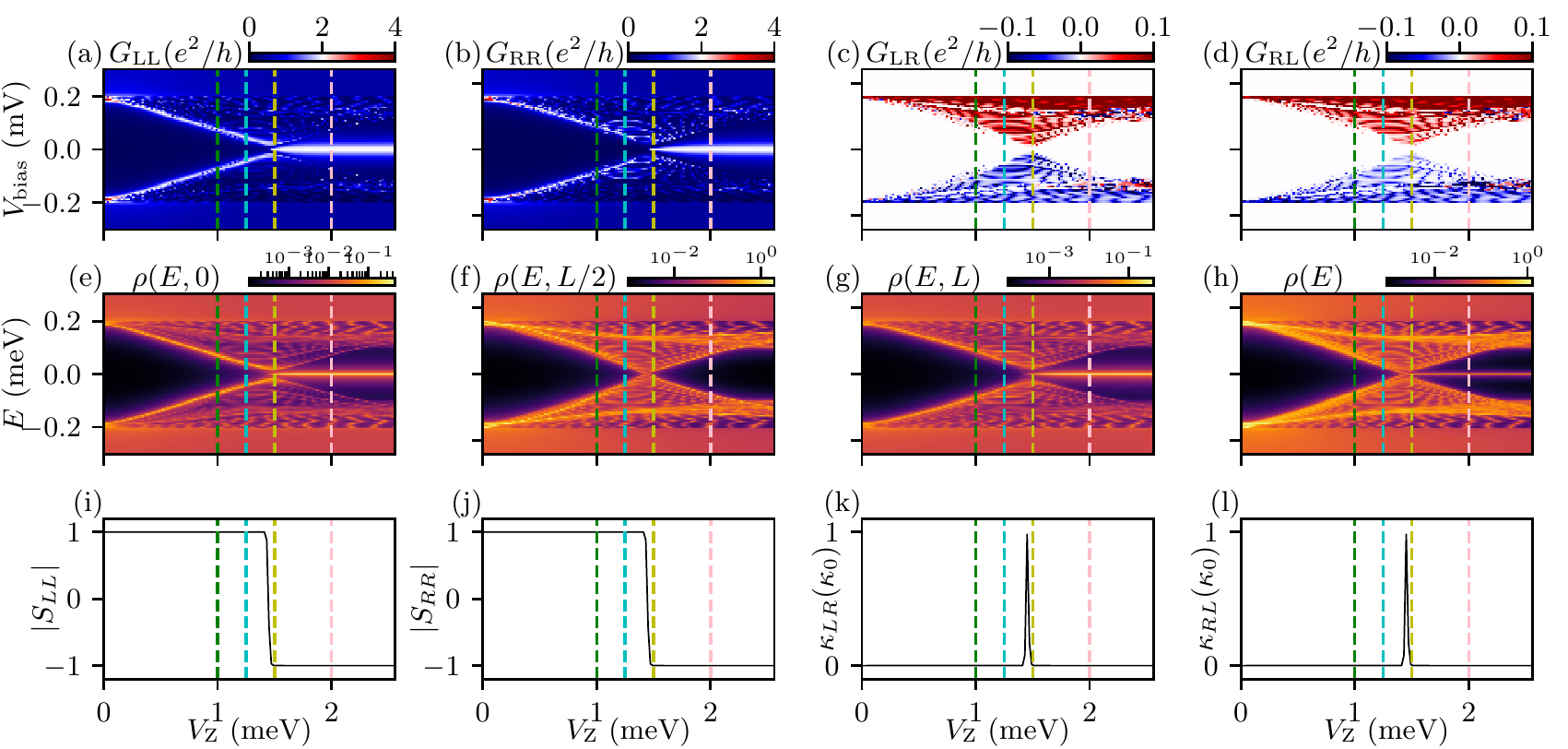}
    \caption{Three-micron wire in the presence of weak disorder ($\sigma=0.3$ meV). 
    (a)-(d) show the local and nonlocal conductances;
    (e)-(h) show the LDOS at $x=0, L/2, L$, and DOS, respectively; Additional LDOS results in the bulk of the wire are presented in Fig.~\ref{fig:L_muVar0.3_3.0_LDOS} (Appendix~\ref{app:LDOS});
    (i)-(l) shows the topological visibility and thermal conductance from both ends. 
    The corresponding wavefunctions and their localization lengths at four typical $V_\text{Z}$ values, as indicated by the colored dashed line, are presented in Fig.~\ref{fig:wf_muVar0.3_3.0} (Appendix~\ref{app:localization}).}
    \label{fig:L_muVar0.3_3.0}
\end{figure*}
With an understanding of the pristine limit, we demonstrate the robustness of the MZMs against weak disorder ($\sigma=0.3$ meV). In Figs.~\ref{fig:L_muVar0.3_1.0} and~\ref{fig:L_muVar0.3_3.0}, we present the same disorder configuration for two different wire lengths, where the longer wire is simply a dilation of the shorter wire, with the spatial disorder profile throughout the wire being the same.

In the presence of weak disorder, Figs.~\ref{fig:L_muVar0.3_1.0}(a)-(b) and Figs.~\ref{fig:L_muVar0.3_3.0}(a)-(b) still show good ZBCPs in the local conductance spectra. Additionally, the gap-closing and reopening features in Figs.~\ref{fig:L_muVar0.3_1.0}(c)-(d) and Figs.~\ref{fig:L_muVar0.3_3.0}(c)-(d), are also manifest, with the only difference being the fact that the minimal gap right at the TQPT is slightly larger than zero due to the effect of weak disorder.
In the LDOS shown in Figs.~\ref{fig:L_muVar0.3_1.0}(e)-(f) and~\ref{fig:L_muVar0.3_3.0}(e)-(f), the low-energy states are localized at the two ends of the wire, and no extra low-energy state is present in the bulk.
The topological visibility and thermal conductances in Figs.~\ref{fig:L_muVar0.3_1.0}(i)-(l) and Figs.~\ref{fig:L_muVar0.3_3.0}(i)-(l) exhibit the consistent features as in the pristine wire, which confirms the topological nature of the lowest state in the presence of weak disorder. 
We note that weak disorder affects neither the qualitative nor the quantitative aspects of the pristine topological physics with Figs.~\ref{fig:L_muVar0.3_1.0} and~\ref{fig:L_muVar0.3_3.0} looking almost identical to Figs.~\ref{fig:L_pris_1.0} and~\ref{fig:L_pris_3.0} although the disorder strength (0.3 meV) is larger than the pristine induced gap (0.2 meV) at zero magnetic field.  This is the robustness or immunity expected for a topological system.


\subsection{Intermediate disorder}\label{sec:intermediate}

\begin{figure*}[htbp]
    \centering
    \includegraphics[width=6.8in]{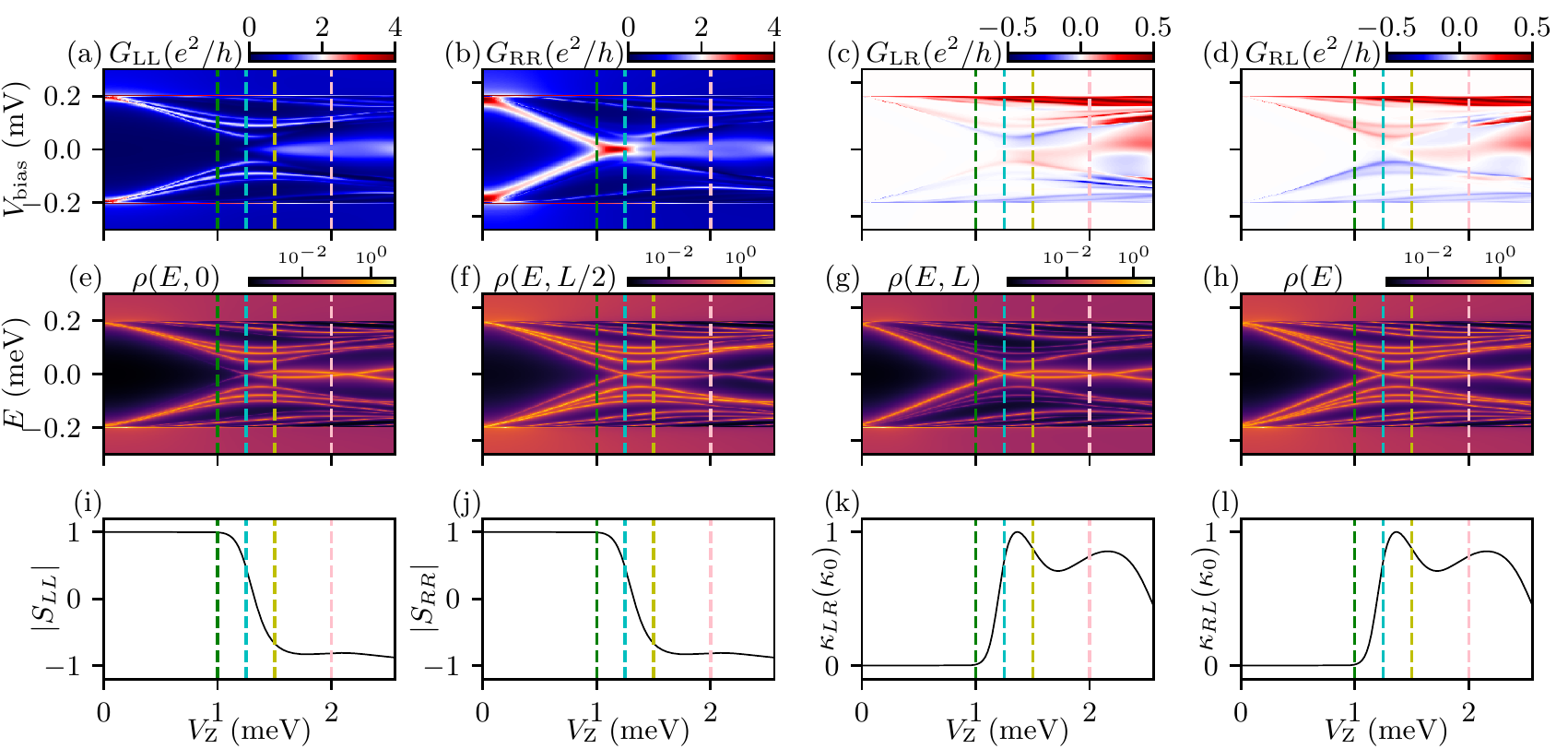}
    \caption{0.5-micron wire in the presence of intermediate disorder ($\sigma=2.5$ meV). 
    (a)-(d) show the local and nonlocal conductances;
    (e)-(h) show the LDOS at $x=0, L/2, L$, and total DOS, respectively; Additional LDOS results in the bulk of the wire are presented in Fig.~\ref{fig:L_muVar2.5_0.5_LDOS} (Appendix~\ref{app:LDOS});
    (i)-(l) shows the topological visibility and thermal conductance from both ends. 
    The corresponding wavefunctions and their localization lengths at four typical $V_\text{Z}$ values, as indicated by the colored dashed line, are presented in Fig.~\ref{fig:wf_muVar2.5_0.5} (Appendix~\ref{app:localization}).}
    \label{fig:L_muVar2.5_0.5}
\end{figure*}

\begin{figure*}[htbp]
    \centering
    \includegraphics[width=6.8in]{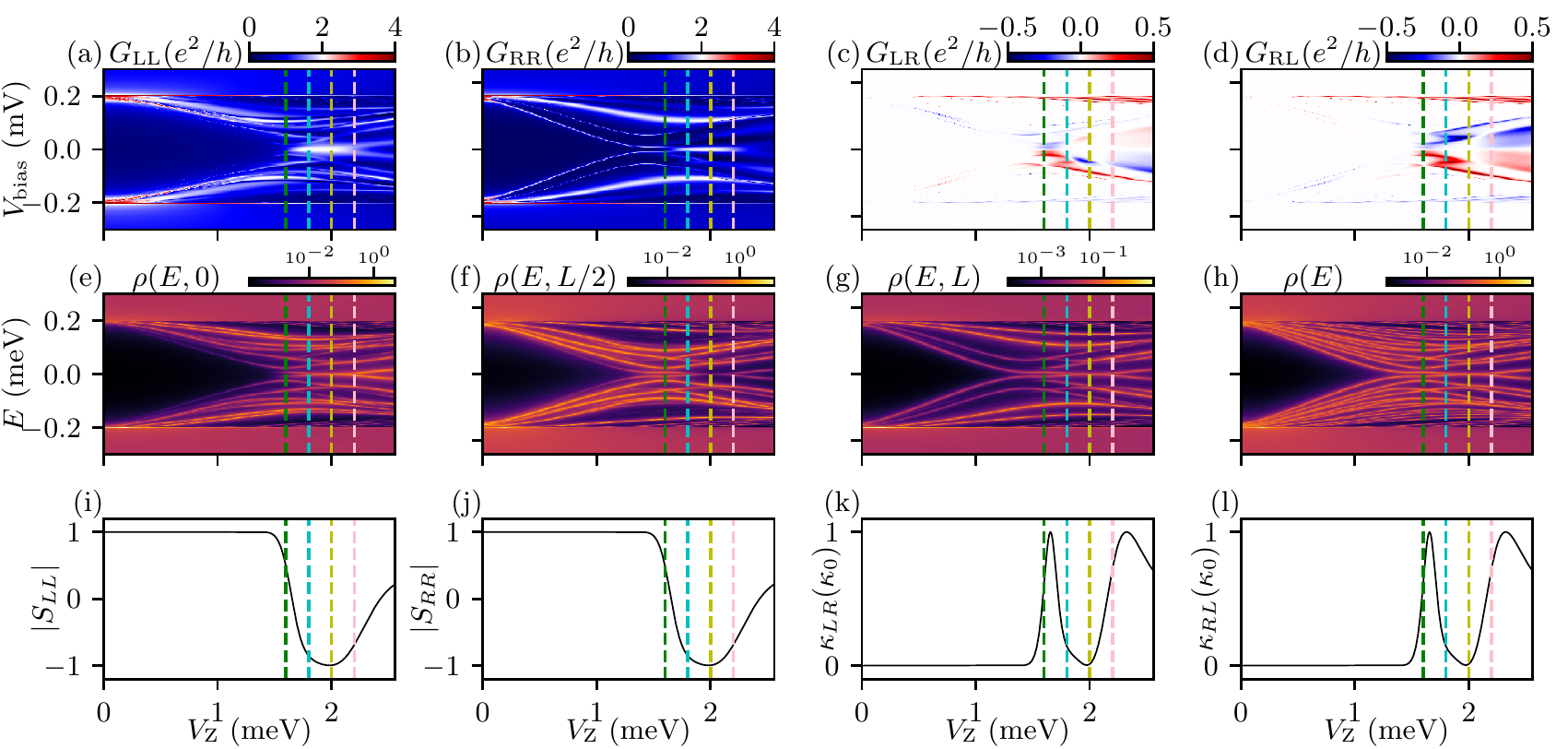}
    \caption{One-micron wire in the presence of intermediate disorder ($\sigma=2.5$ meV). 
    (a)-(d) show the local and nonlocal conductances;
    (e)-(h) show the LDOS at $x=0, L/2, L$, and total DOS, respectively; Additional LDOS results in the bulk of the wire are presented in Fig.~\ref{fig:L_muVar2.5_1.0_LDOS} (Appendix~\ref{app:LDOS});
    (i)-(l) shows the topological visibility and thermal conductance from both ends. 
    The corresponding wavefunctions and their localization lengths at four typical $V_\text{Z}$ values, as indicated by the colored dashed line, are presented in Fig.~\ref{fig:wf_muVar2.5_1.0} (Appendix~\ref{app:localization}).}
    \label{fig:L_muVar2.5_1.0}
\end{figure*}

\begin{figure*}[htbp]
    \centering
    \includegraphics[width=6.8in]{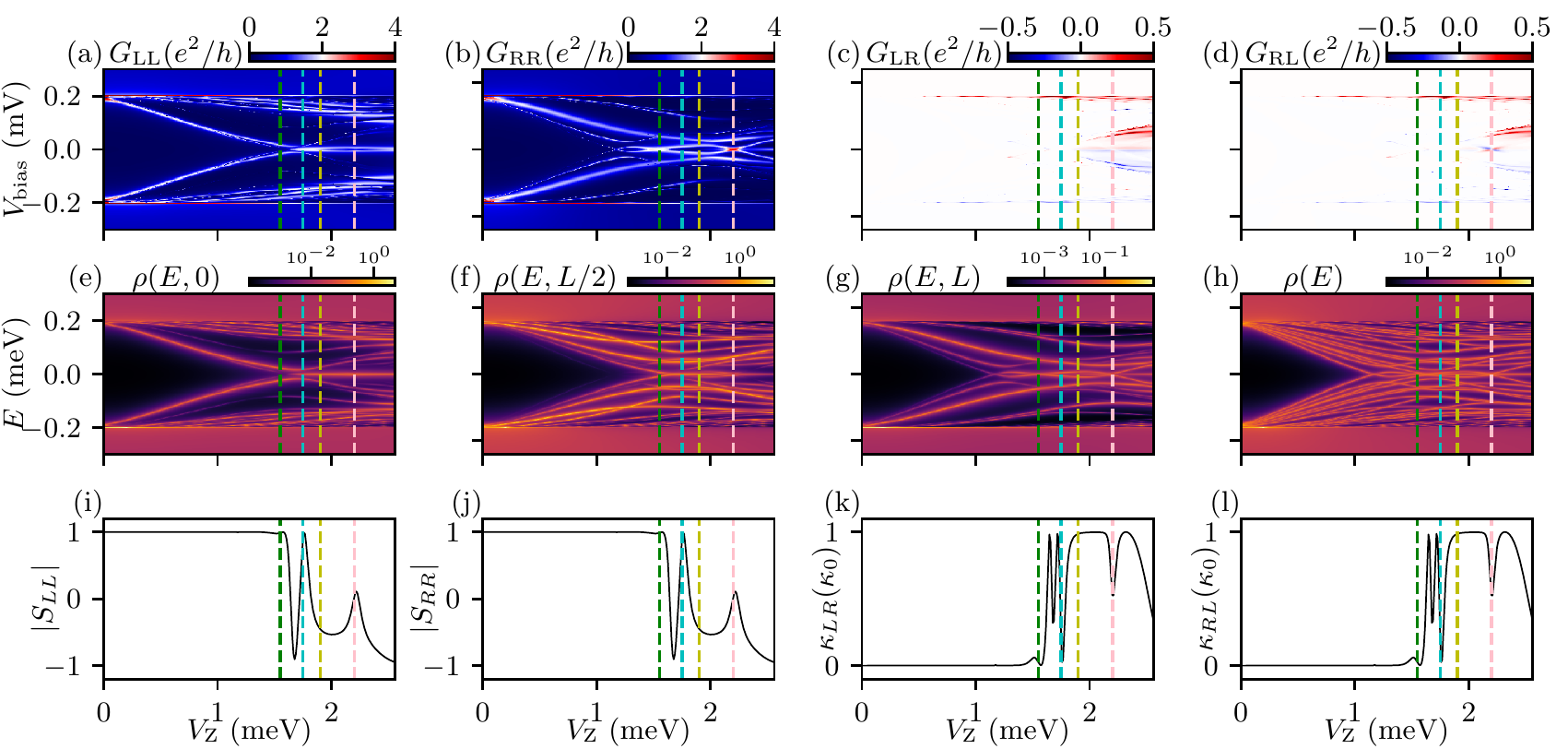}
    \caption{1.5-micron wire in the presence of intermediate disorder ($\sigma=2.5$ meV). 
    (a)-(d) show the local and nonlocal conductances;
    (e)-(h) show the LDOS at $x=0, L/2, L$, and total DOS, respectively; Additional LDOS results in the bulk of the wire are presented in Fig.~\ref{fig:L_muVar2.5_1.5_LDOS} (Appendix~\ref{app:LDOS});
    (i)-(l) shows the topological visibility and thermal conductance from both ends. 
    The corresponding wavefunctions and their localization lengths at four typical $V_\text{Z}$ values, as indicated by the colored dashed line, are presented in Fig.~\ref{fig:wf_muVar2.5_1.5} (Appendix~\ref{app:localization}).}
    \label{fig:L_muVar2.5_1.5}
\end{figure*}

\begin{figure*}[htbp]
    \centering
    \includegraphics[width=6.8in]{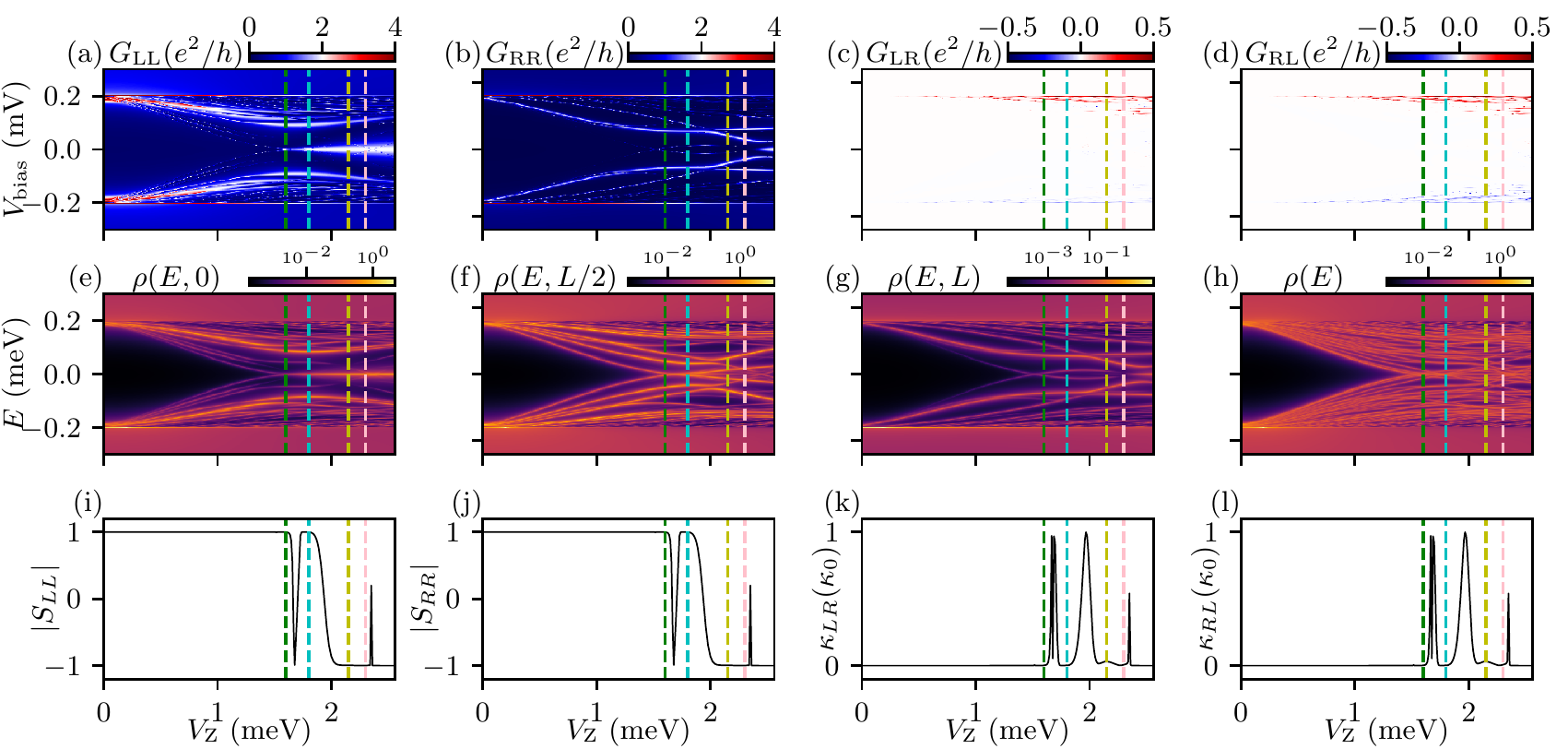}
    \caption{Three-micron wire in the presence of intermediate disorder ($\sigma=2.5$ meV). 
    (a)-(d) show the local and nonlocal conductances;
    (e)-(h) show the LDOS at $x=0, L/2, L$, and total DOS, respectively; Additional LDOS results in the bulk of the wire are presented in Fig.~\ref{fig:L_muVar2.5_3.0_LDOS} (Appendix~\ref{app:LDOS});
    (i)-(l) shows the topological visibility and thermal conductance from both ends. 
    The corresponding wavefunctions and their localization lengths at four typical $V_\text{Z}$ values, as indicated by the colored dashed line, are presented in Fig.~\ref{fig:wf_muVar2.5_3.0} (Appendix~\ref{app:localization}).}
    \label{fig:L_muVar2.5_3.0}
\end{figure*}

\begin{figure*}[htbp]
    \centering
    \includegraphics[width=6.8in]{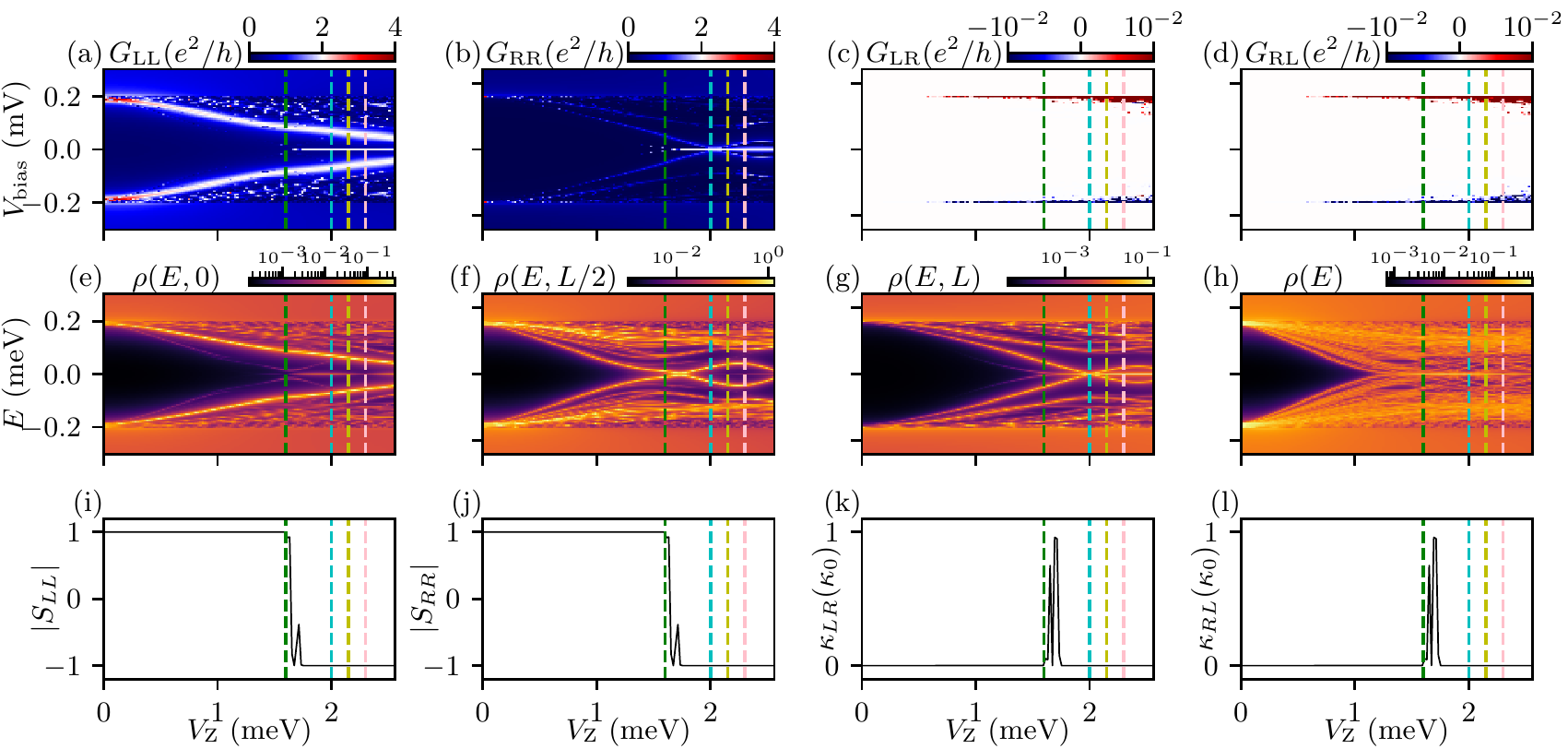}
    \caption{Ten-micron wire in the presence of intermediate disorder ($\sigma=2.5$ meV). 
    (a)-(d) show the local and nonlocal conductances;
    (e)-(h) show the LDOS at $x=0, L/2, L$, and total DOS, respectively; Additional LDOS results in the bulk of the wire are presented in Fig.~\ref{fig:L_muVar2.5_10.0_LDOS} (Appendix~\ref{app:LDOS});
    (i)-(l) shows the topological visibility and thermal conductance from both ends. 
    The corresponding wavefunctions and their localization lengths at four typical $V_\text{Z}$ values, as indicated by the colored dashed line, are presented in Fig.~\ref{fig:wf_muVar2.5_10.0} (Appendix~\ref{app:localization}).}
    \label{fig:L_muVar2.5_10.0}
\end{figure*}

\begin{figure*}[htbp]
    \centering
    \includegraphics[width=6.8in]{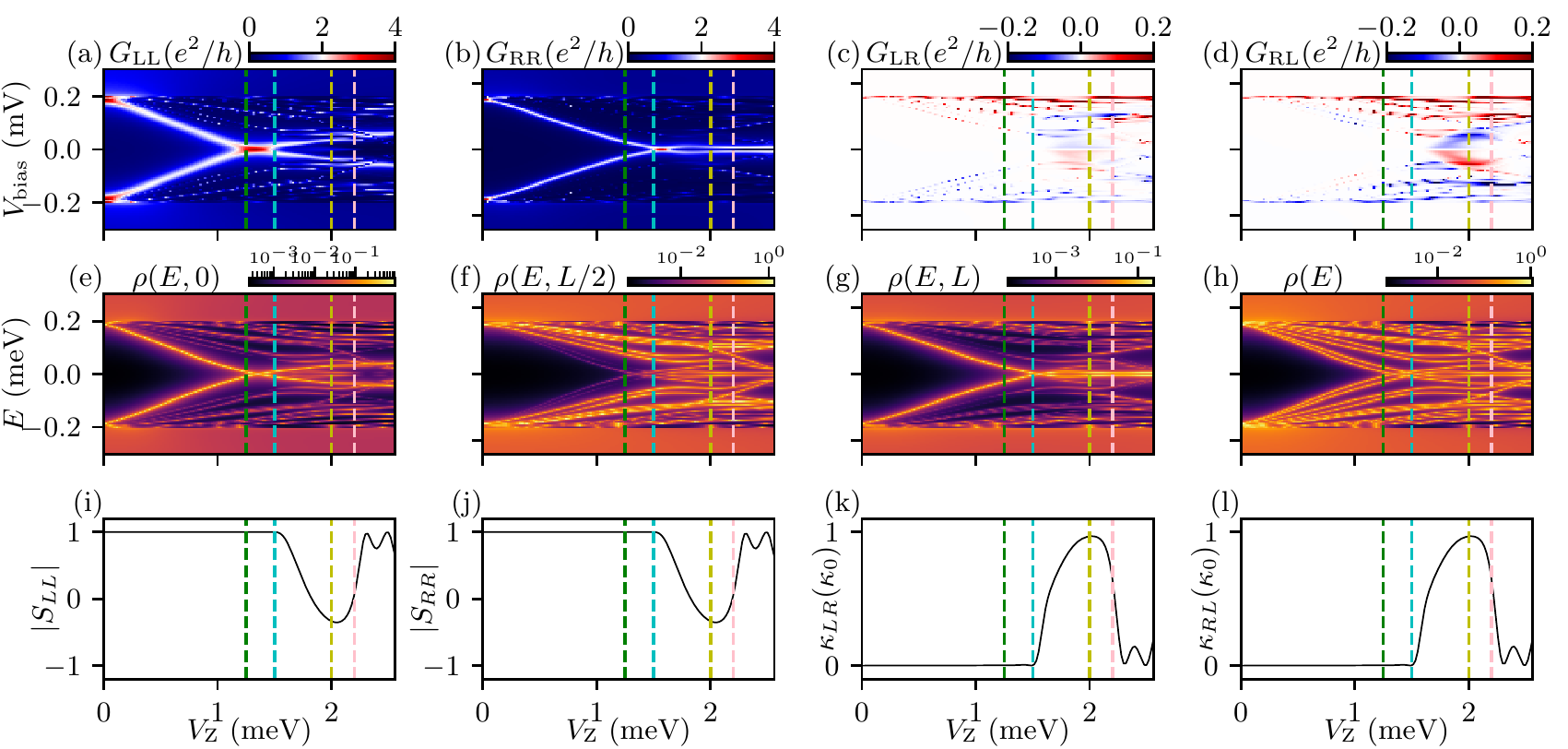}
    \caption{One-micron wire in the presence of intermediate disorder ($\sigma=3$ meV). 
    (a)-(d) show the local and nonlocal conductances;
    (e)-(h) show the LDOS at $x=0, L/2, L$, and total DOS, respectively; Additional LDOS results in the bulk of the wire are presented in Fig.~\ref{fig:L_muVar3.0_1.0_LDOS} (Appendix~\ref{app:LDOS});
    (i)-(l) shows the topological visibility and thermal conductance from both ends. 
    The corresponding wavefunctions and their localization lengths at four typical $V_\text{Z}$ values, as indicated by the colored dashed line, are presented in Fig.~\ref{fig:wf_muVar3.0_1.0} (Appendix~\ref{app:localization}).}
    \label{fig:L_muVar3.0_1.0}
\end{figure*}

\begin{figure*}[htbp]
    \centering
    \includegraphics[width=6.8in]{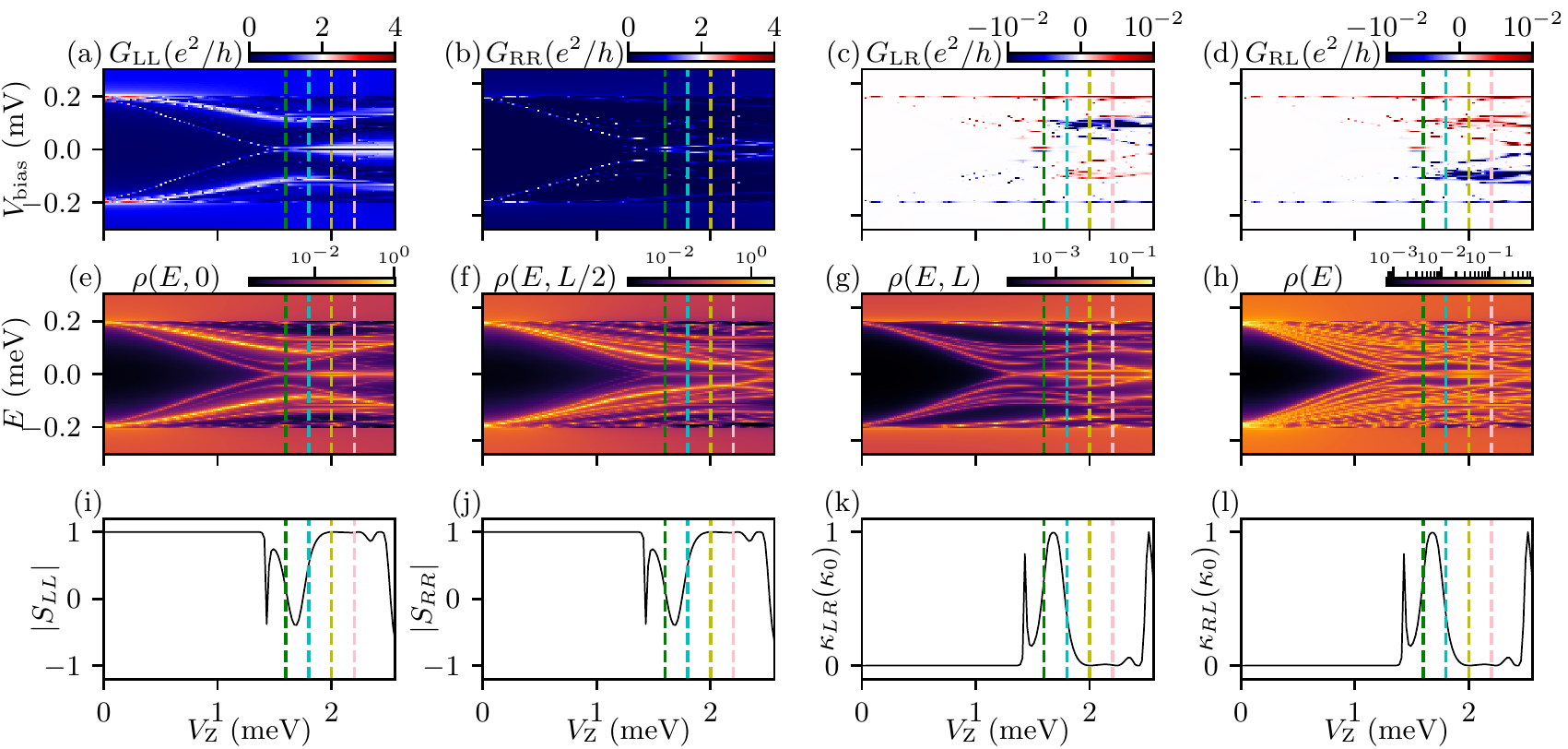}
    \caption{Two-micron wire in the presence of intermediate disorder ($\sigma=3$ meV). 
    (a)-(d) show the local and nonlocal conductances;
    (e)-(h) show the LDOS at $x=0, L/2, L$, and total DOS, respectively; Additional LDOS results in the bulk of the wire are presented in Fig.~\ref{fig:L_muVar3.0_2.0_LDOS} (Appendix~\ref{app:LDOS});
    (i)-(l) shows the topological visibility and thermal conductance from both ends. 
    The corresponding wavefunctions and their localization lengths at four typical $V_\text{Z}$ values, as indicated by the colored dashed line, are presented in Fig.~\ref{fig:wf_muVar3.0_2.0} (Appendix~\ref{app:localization}).}
    \label{fig:L_muVar3.0_2.0}
\end{figure*}

\begin{figure*}[htbp]
    \centering
    \includegraphics[width=6.8in]{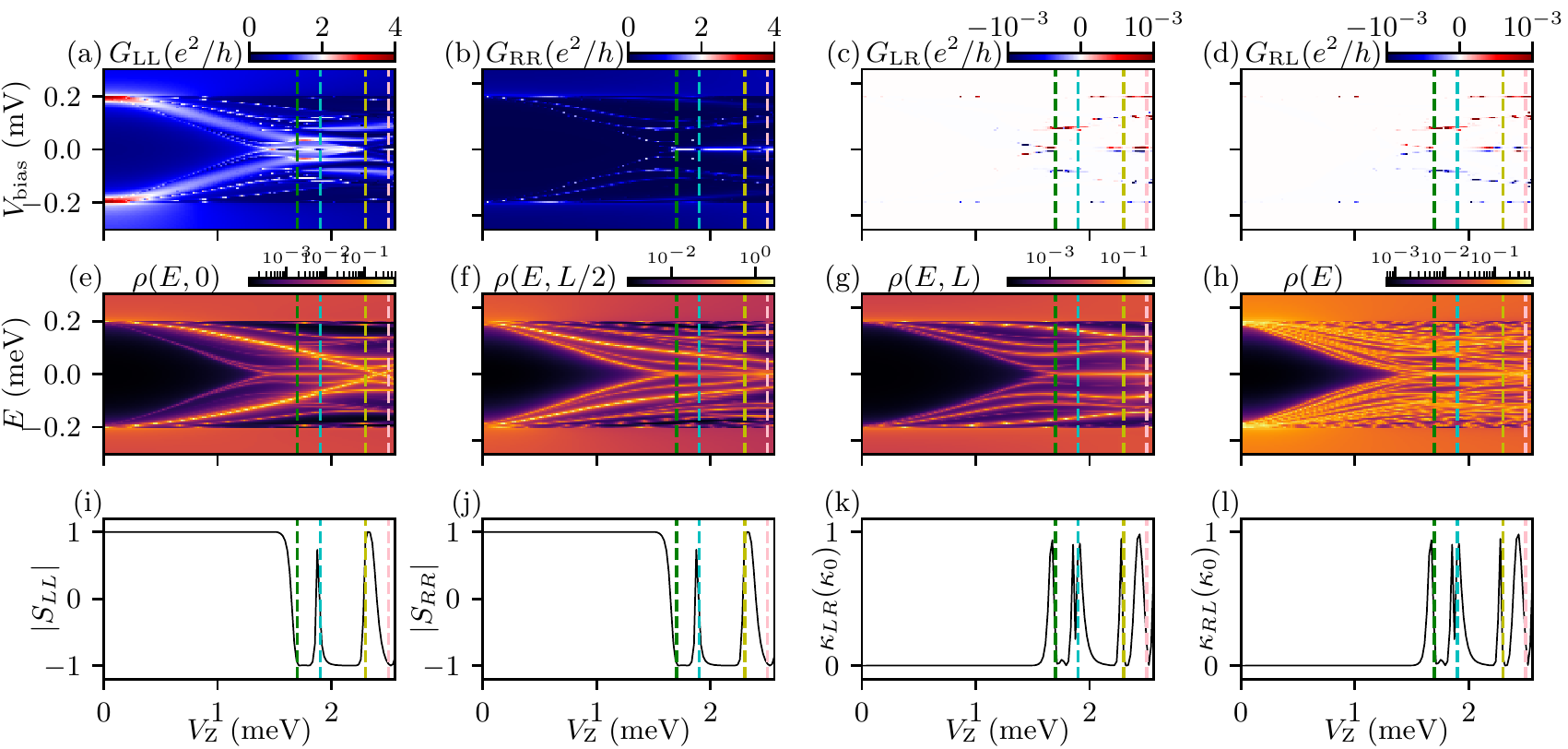}
    \caption{Three-micron wire in the presence of intermediate disorder ($\sigma=3$ meV). 
    (a)-(d) show the local and nonlocal conductances;
    (e)-(h) show the LDOS at $x=0, L/2, L$, and total DOS, respectively; Additional LDOS results in the bulk of the wire are presented in Fig.~\ref{fig:L_muVar3.0_3.0_LDOS} (Appendix~\ref{app:LDOS});
    (i)-(l) shows the topological visibility and thermal conductance from both ends. 
    The corresponding wavefunctions and their localization lengths at four typical $V_\text{Z}$ values, as indicated by the colored dashed line, are presented in Fig.~\ref{fig:wf_muVar3.0_3.0} (Appendix~\ref{app:localization}).}
    \label{fig:L_muVar3.0_3.0}
\end{figure*}

Although MZMs are robust in the presence of weak disorder, the situation becomes more complex as the strength of disorder increases. 
In this section, we focus on the intermediate level of disorder to show that the finite nanowire exhibits intricate behaviors that are highly dependent on details such as disorder configurations and wire length. 
Specifically, we choose two distinct disorder configurations (Figs.~\ref{fig:L_muVar2.5_0.5}-~\ref{fig:L_muVar2.5_10.0} and Figs.~\ref{fig:L_muVar3.0_1.0}-~\ref{fig:L_muVar3.0_3.0}), varying the wire length from extremely short ($L= 0.5$ microns) to very long ($L= 10$ microns). 
We believe that the Microsoft samples are in the intermediate disorder regime, but we do not know the experimental disorder configuration and strength, making it difficult to say anything definitive, quantitatively.

We start with the extremely short wire ($L=0.5$ micron) as shown in Fig.~\ref{fig:L_muVar2.5_0.5}.
We find that a ZBCP, which may even be correlated from both ends, can appear in the local conductance spectrum below the putative TQPT at $V_\text{Z}=1.41$ meV (Fig.~\ref{fig:L_muVar2.5_0.5}(b)) with considerably higher conductance above $2e^2/h$, which is a characteristic feature of disorder-induced ZBCPs~\cite{pan2020physical}. This is not the usual MZM-induced ZBCP which cannot exceed $2e^2/h$. Beyond the TQPT at $V_\text{Z}=1.41$ meV, however, the local conductance spectra from both ends show ZBCP quantized at $2e^2/h$ corresponding to good ZBCPs.
In the nonlocal conductance spectra, as shown in Figs.~\ref{fig:L_muVar2.5_0.5}(c)-(d), we observe a clear gap-closing feature but no gap reopening even though the topological visibility indicates a nontrivial topology beyond $V_\text{Z}=1.4$ meV in Figs.~\ref{fig:L_muVar2.5_0.5}(i)-~\ref{fig:L_muVar2.5_0.5}(j), and the first excited state inside the continuum spectrum is gapped from the zero energy MZM as shown in the total DOS in Fig.~\ref{fig:L_muVar2.5_0.5}(h).
The thermal conductance manifests a weak TQPT at best.
The absence of gap-reopening is due to the significant finite size effect in such a short wire, leading to a large overlap of MZMs from both ends of the wire, as shown in Fig.~\ref{fig:wf_muVar2.5_0.5} (Appendix~\ref{app:LDOS}).
Furthermore, this large overlap is a result of the MZMs extending to the bulk regime of the short wire, which is also confirmed by the LDOS, where Majorana oscillations are apparent even in the middle of the wire. We find that in addition to the localized states at two wire ends as shown in Figs.~\ref{fig:L_muVar2.5_0.5}(e) and~\ref{fig:L_muVar2.5_0.5}(g), the LDOS in the middle of the wire, $\rho(E,L/2)$, (Fig.~\ref{fig:L_muVar2.5_0.5}(h)) is also finite near zero energy.
Therefore, the significant finite size effect makes the usual diagnostics for topology confusing in short wires, i.e., the phase transition in the topological visibility not being sharp, and thermal conductances not vanishing beyond the TQPT at around $V_\text{Z}=1.4$ meV (Figs.~\ref{fig:L_muVar2.5_0.5}(k) and~\ref{fig:L_muVar2.5_0.5}(l)).

As the nanowire length increases, the finite size effect becomes less severe, and the phase transition becomes sharper, as shown in Figs.~\ref{fig:L_muVar2.5_1.0}(i) and~\ref{fig:L_muVar2.5_1.0}(j) for a one-micron wire.
However, the gap reopening feature remains absent, as shown in Figs.~\ref{fig:L_muVar2.5_1.0}(c)-(d). 
The gap size beyond TQPT is very small, and it does not increase as the Zeeman field increases. This feature is similar to the results from the recent Microsoft experiment~\cite{microsoftquantum2023inasal}.
The minimal topological gap provides little protection to the MZMs, rendering them susceptible to local disorder in the bulk of the wire. 
We also note that the LDOS near zero energy is finite throughout the wire as shown in Fig.~\ref{fig:L_muVar2.5_1.0}(e)-(g) in contrast to the pristine and weak disorder limit where the LDOS near zero energy has a finite value only at wire ends. 
This implies a considerable amount of disorder-induced near-zero energy subgap fermionic Andreev bound states throughout the wire, considerably suppressing topological physics as well as interpretation of data because of considerable sample-to-sample variations arising from the specific mesoscopic disorder configuration.

\begin{figure}[htbp]
    \centering
    \includegraphics[width=3.4in]{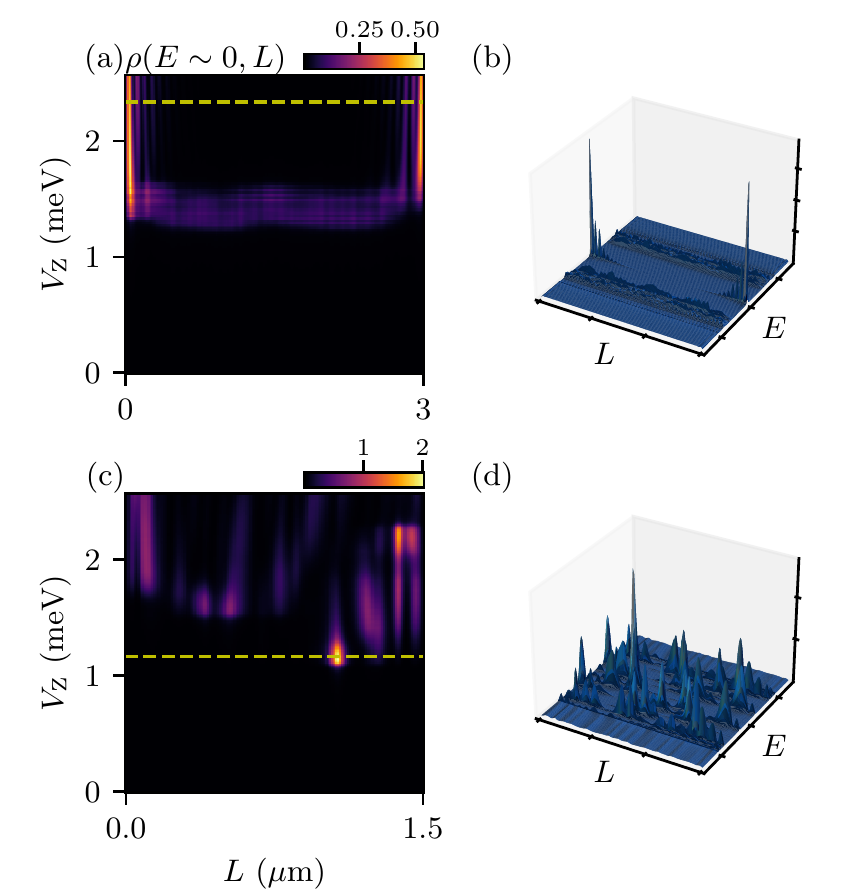}
    \caption{Comparison of LDOS for weak disorder (top panel) and intermediate disorder (bottom panel). (a,c) show the LDOS at different $V_\text{Z}$ near zero energy for the two cases; (b,d) show LDOS at a specific $V_\text{Z}$ indicated by the dashed line in (a,c). The parameters in (a,b) correspond to Fig.~\ref{fig:L_muVar0.3_3.0} and the parameters in (c,d) correspond to Fig.~\ref{fig:L_muVar2.5_1.5}.}
    \label{fig:LDOS_peak}
\end{figure}

For an even longer wire, such as a 1.5-micron wire in Fig.~\ref{fig:L_muVar2.5_1.5} (see also Fig.~\ref{fig:L_muVar2.5_3.0} in this context), an important feature to note is that while we do find an operational topological gap beyond the TQPT in the nonlocal conductance, as shown in Figs.~\ref{fig:L_muVar2.5_1.5}(c) and~\ref{fig:L_muVar2.5_1.5}(d), we do not observe the existence of isolated MZM due to the Griffiths phase physics~\cite{motrunich2001griffiths} of disorder-induced low-energy DOS throughout the wire (and not just at the ends), implying multiple MZMs along the wire in the bulk. 
[This is even more evident in a direct comparison between a weak disorder case (top panel in Fig.~\ref{fig:LDOS_peak}) and an intermediate disorder case (bottom panel in Fig.~\ref{fig:LDOS_peak}).
In Figs.~\ref{fig:LDOS_peak}(a,c), we plot the average LDOS near zero energy as a function of Zeeman field, and find that weak disorder only shows the localization of low-energy states at the wire ends, while the intermediate disorder can host multiple low-energy states localized in the bulk of the wire.]
This implies that, in the presence of intermediate disorder, even if a fragile topological gap is present, as shown in Figs.~\ref{fig:L_muVar2.5_1.5}(c)-(d), there could be too many low-energy states throughout the wire as shown in Fig.~\ref{fig:LDOS_peak}(d). In contrast, the low-energy state in the presence of weak disorder only has two peaks localized at the wire ends as shown in Fig.~\ref{fig:LDOS_peak}(b). 
In addition, the fluctuations of the topological visibility in Figs.~\ref{fig:L_muVar2.5_1.5}(i) and~\ref{fig:L_muVar2.5_1.5}(j) between +1 and -1 indicate the multiple effective transitions between topological and nontopological phases as Zeeman field increases. 
This situation is qualitatively similar to what is observed in the Microsoft experiment, where small isolated patches of topological gaps are reported in the parameter space, implying multiple TQPTs.

For an extremely long wire, such as a ten-micron wire shown in Fig.~\ref{fig:L_muVar2.5_10.0}, we observe the appearance of ZBCPs in the local conductance spectra, as shown in Figs.~\ref{fig:L_muVar2.5_10.0}(a)-(b) without evidence of gap closing and reopening due to the disorder. In this long wire limit, intermediate disorder effectively breaks the nanowire into several segments, making the nonlocal transmission between the leads almost impossible.
The basic problem is that the two ends are now more distant from each other than the disorder-induced mean free path, making it impossible for nonlocal conductance to manifest itself.
Similarly, we also find a plethora of low-energy states throughout the wire as shown in Fig.~\ref{fig:L_muVar2.5_10.0}(h) in the total DOS (and in Fig.~\ref{fig:L_muVar2.5_10.0_LDOS} of Appendix~\ref{app:LDOS}).
Nevertheless, the topological visibility in Figs.~\ref{fig:L_muVar2.5_10.0}(i) and~\ref{fig:L_muVar2.5_10.0}(j), and the thermal conductance in Figs.~\ref{fig:L_muVar2.5_10.0}(k) and~\ref{fig:L_muVar2.5_10.0}(l) still indicate one sharp TQPT at $V_\text{Z}=1.41$ meV.
This is an important point: Even when the system may be topological, as implied by the topological invariants, the nonlocal conductance may not manifest any typological gap if the mean free path is shorter than the nominal wire length in disordered systems.  Thus, the measured topological gap in the nonlocal conductance becomes increasingly meaningless in longer wires.

In addition to the first intermediate disorder configuration, we also present a second disorder configuration in the intermediate regime with a slightly stronger disorder strength of $\sigma$ = 3 meV and wire lengths ranging from one micron (Fig.~\ref{fig:L_muVar3.0_1.0}) to three microns (Fig.~\ref{fig:L_muVar3.0_3.0}). 
However, the key features are consistent with the results in Figs.~\ref{fig:L_muVar2.5_0.5}-\ref{fig:L_muVar2.5_10.0}, with a fragile topological gap in nonlocal conductance and numerous low-energy states throughout the wire. These results suggest that the intermediate disorder regime is complex and can lead to multiple effective transitions between topological and nontopological phases as system parameters change.
We believe that this complex situation applies qualitatively to the current Majorana experiments.  In particular, multiple TQPTs in the parameter space with topological and nontopological patches residing close by are observed in the Majorana experiment.  Our results indicate that multiple TQPTs with an increasing magnetic field may be generic in the intermediate disorder regime.

\begin{figure*}[htbp]
    \centering
    \includegraphics[width=6.8in]{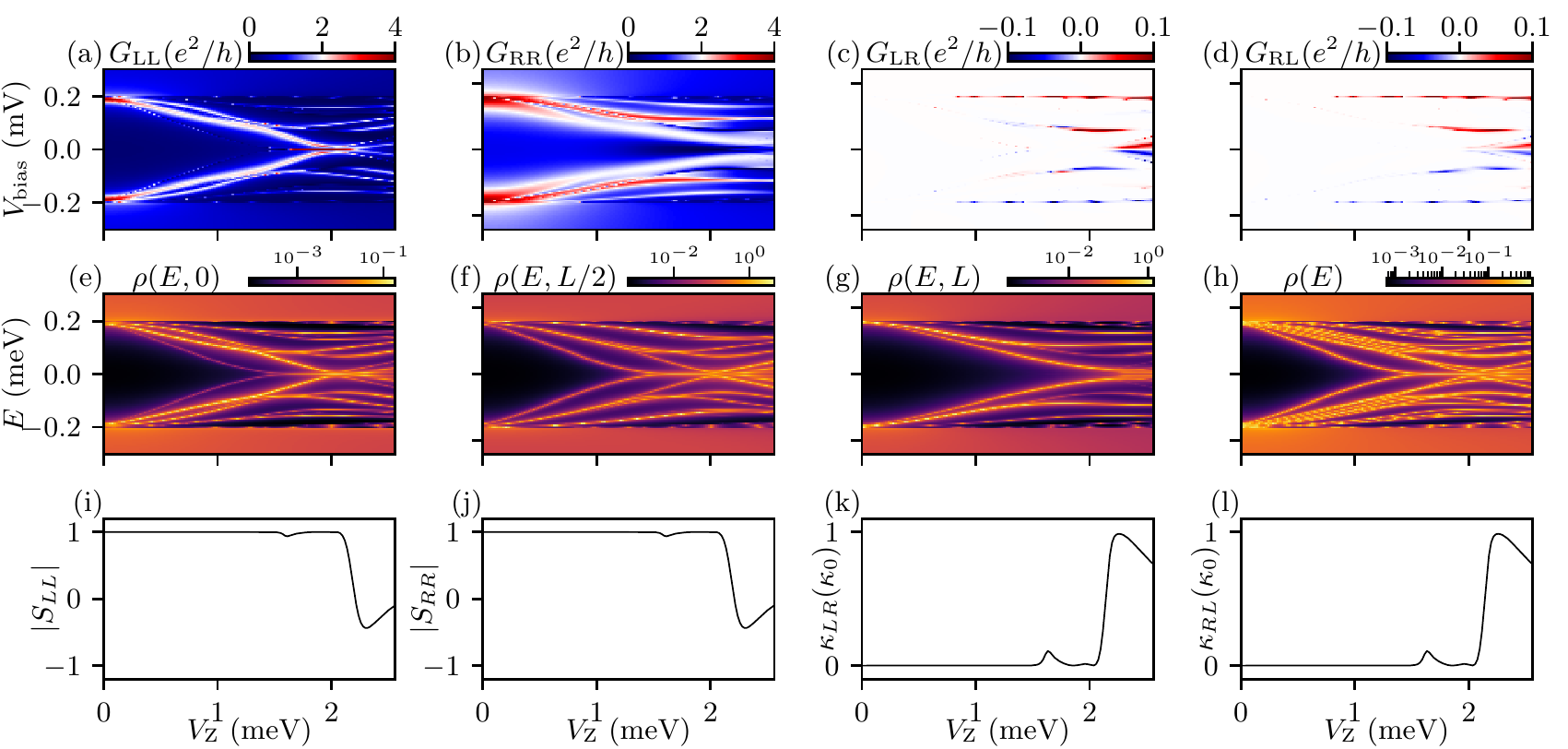}
    \caption{One-micron wire in the presence of intermediate disorder ($\sigma=5$ meV). 
    (a)-(d) show the local and nonlocal conductances;
    (e)-(h) show the LDOS at $x=0, L/2, L$, and total DOS, respectively; Additional LDOS results in the bulk of the wire are presented in Fig.~\ref{fig:L_muVar5.0_1.0_LDOS} (Appendix~\ref{app:LDOS});
    (i)-(l) shows the topological visibility and thermal conductance from both ends. 
    }
    \label{fig:L_muVar5.0_1.0}
\end{figure*}

\begin{figure*}[htbp]
    \centering
    \includegraphics[width=6.8in]{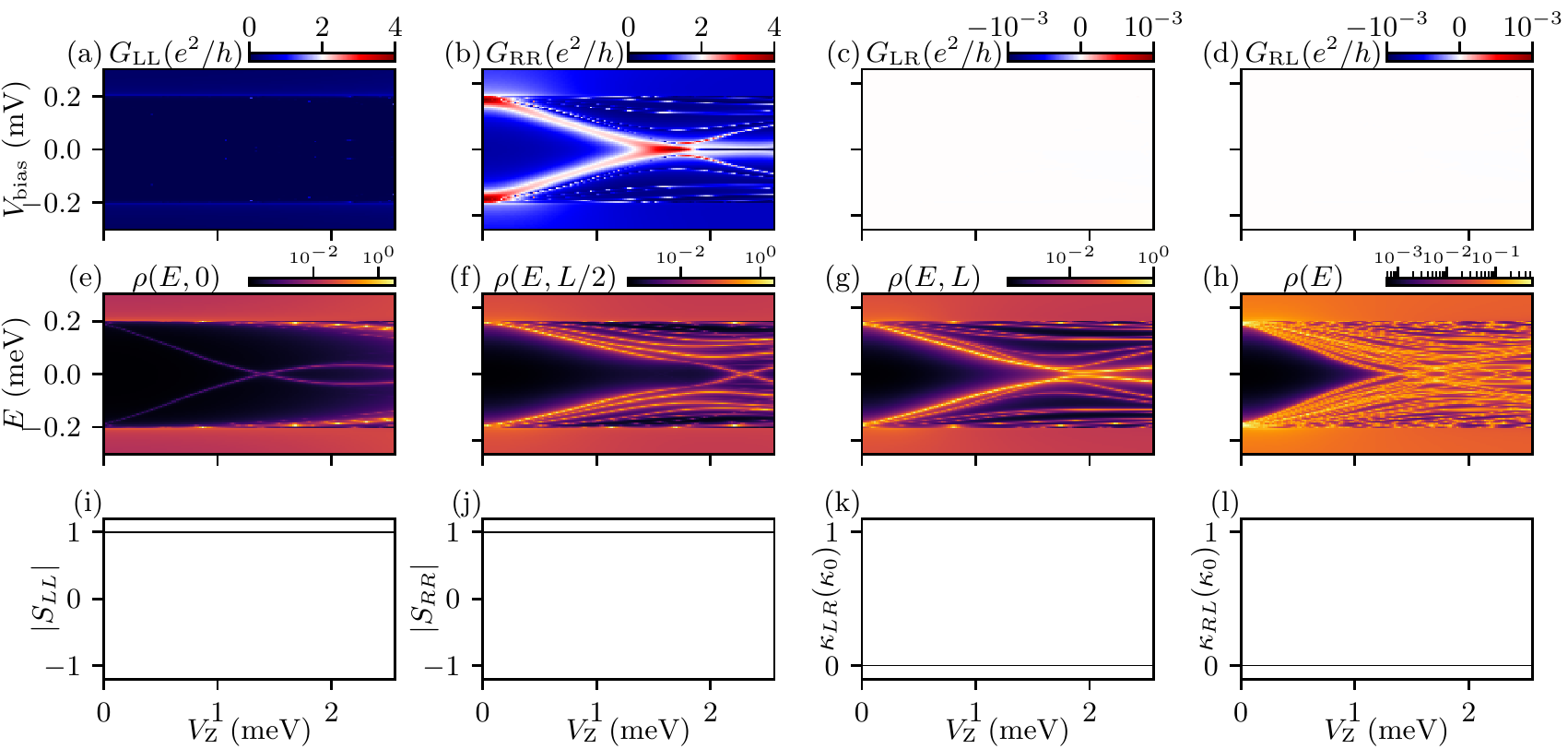}
    \caption{Three-micron wire in the presence of intermediate disorder ($\sigma=5$ meV). 
    (a)-(d) show the local and nonlocal conductances;
    (e)-(h) show the LDOS at $x=0, L/2, L$, and total DOS, respectively; Additional LDOS results in the bulk of the wire are presented in Fig.~\ref{fig:L_muVar5.0_3.0_LDOS} (Appendix~\ref{app:LDOS});
    (i)-(l) shows the topological visibility and thermal conductance from both ends. 
    }
    \label{fig:L_muVar5.0_3.0}
\end{figure*}

\begin{figure*}[htbp]
    \centering
    \includegraphics[width=6.8in]{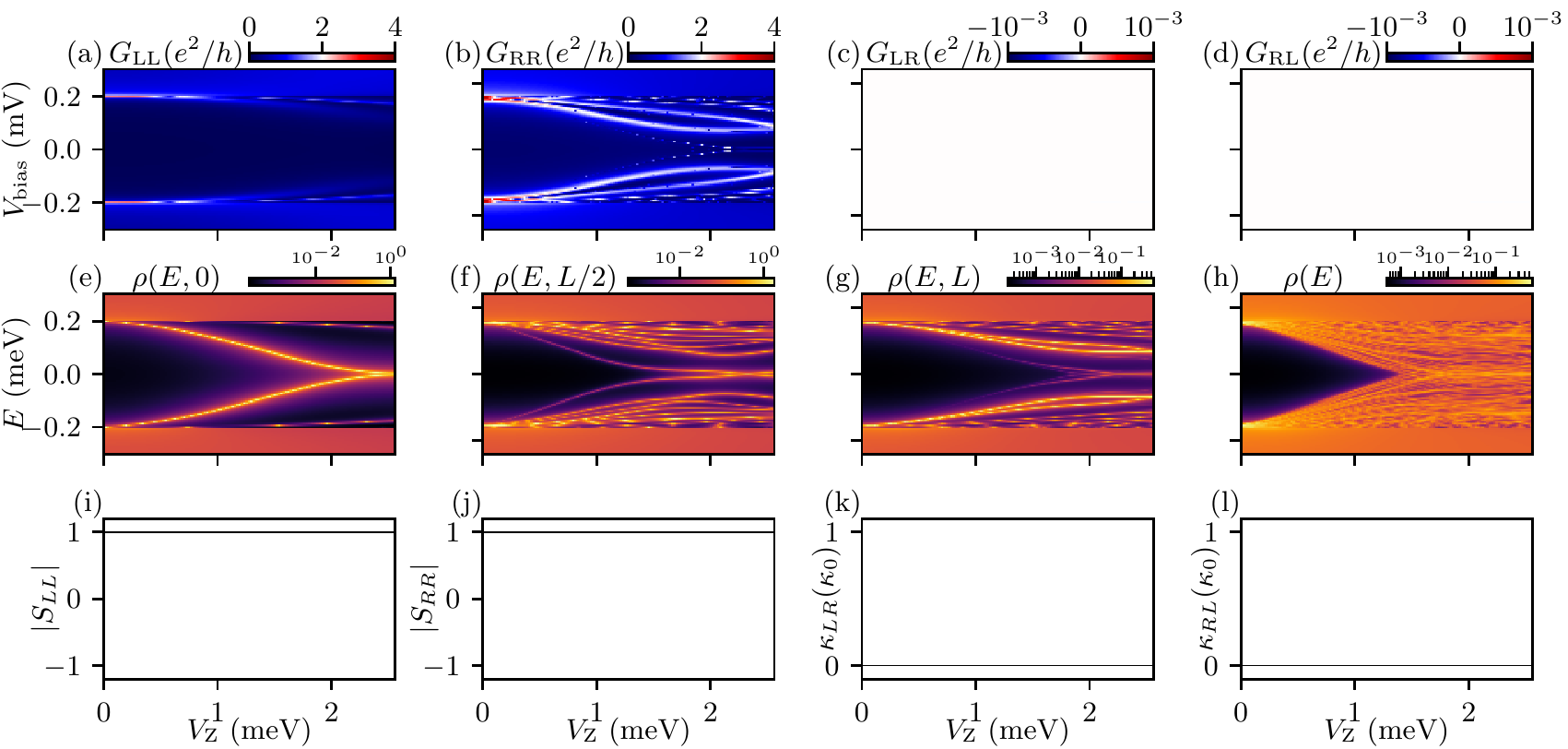}
    \caption{Ten-micron wire in the presence of intermediate disorder ($\sigma=5$ meV). 
    (a)-(d) show the local and nonlocal conductances;
    (e)-(h) show the LDOS at $x=0, L/2, L$, and total DOS, respectively; Additional LDOS results in the bulk of the wire are presented in Fig.~\ref{fig:L_muVar5.0_10.0_LDOS} (Appendix~\ref{app:LDOS});
    (i)-(l) shows the topological visibility and thermal conductance from both ends. 
    }
    \label{fig:L_muVar5.0_10.0}
\end{figure*}

\begin{figure*}[htbp]
    \centering
    \includegraphics[width=6.8in]{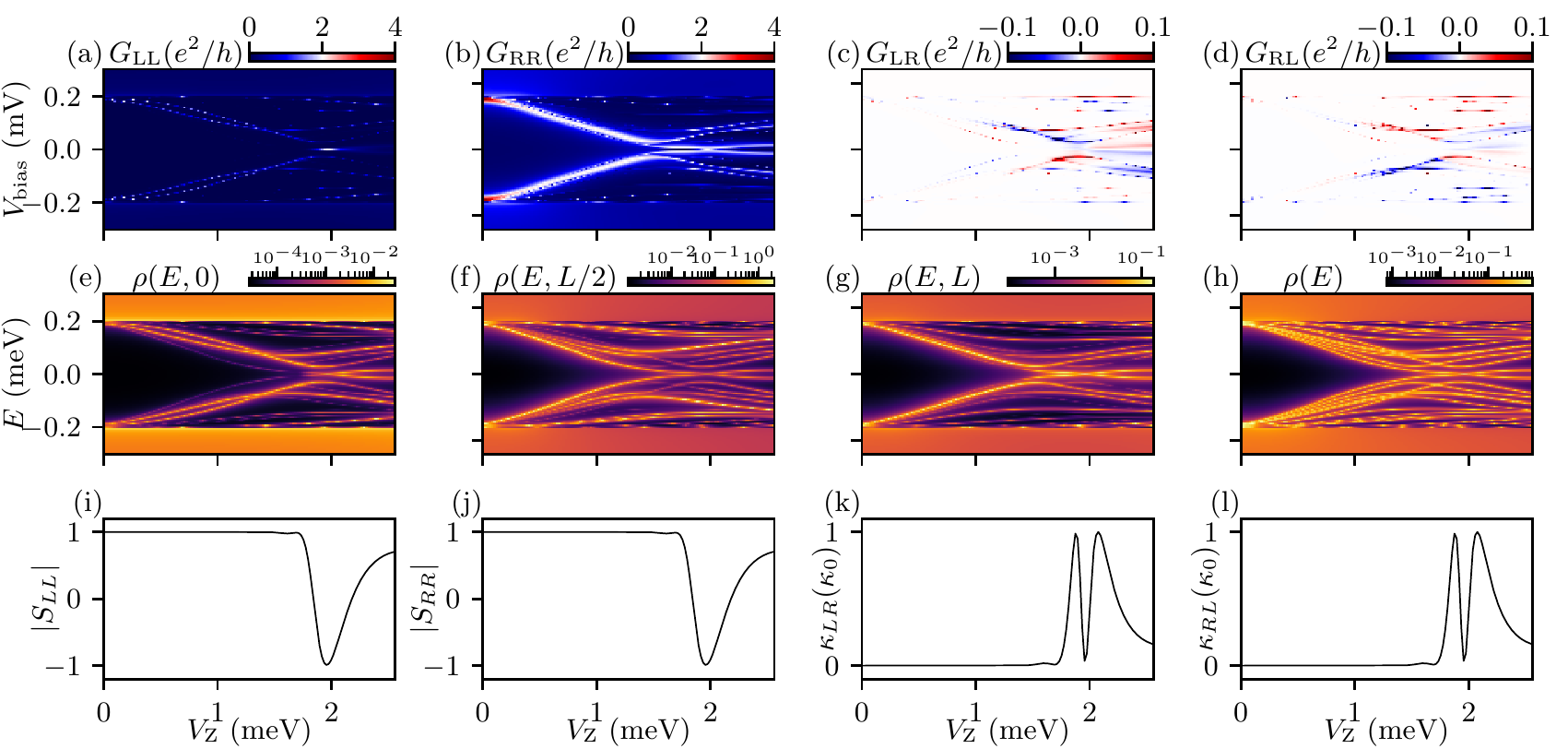}
    \caption{One-micron wire in the presence of intermediate disorder ($\sigma=5$ meV). 
    (a)-(d) show the local and nonlocal conductances;
    (e)-(h) show the LDOS at $x=0, L/2, L$, and total DOS, respectively; Additional LDOS results in the bulk of the wire are presented in Fig.~\ref{fig:L_muVar5.0_1.0_2_LDOS} (Appendix~\ref{app:LDOS});
    (i)-(l) shows the topological visibility and thermal conductance from both ends. 
    }
    \label{fig:L_muVar5.0_1.0_2}
\end{figure*}

\begin{figure*}[htbp]
    \centering
    \includegraphics[width=6.8in]{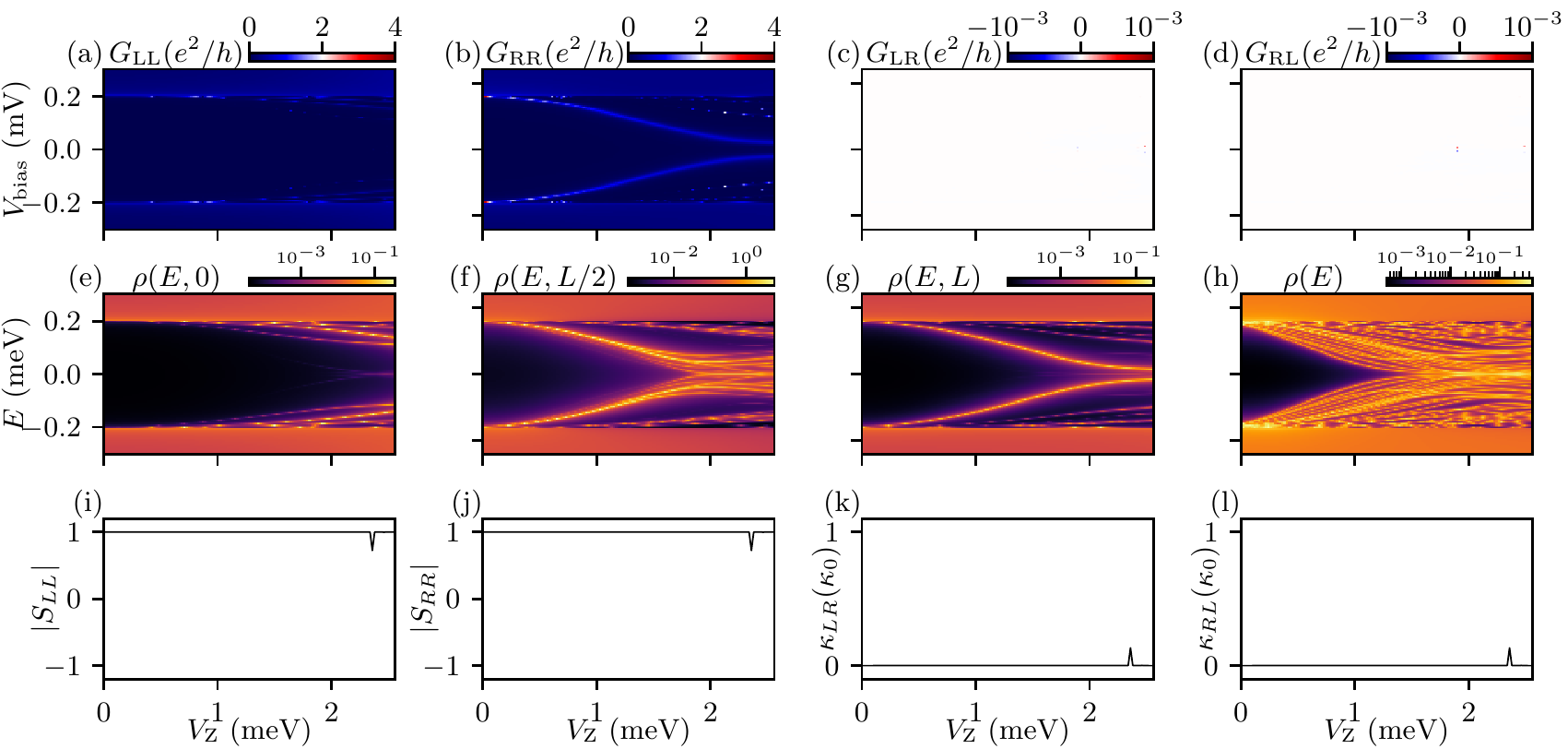}
    \caption{Three-micron wire in the presence of intermediate disorder ($\sigma=5$ meV). 
    (a)-(d) show the local and nonlocal conductances;
    (e)-(h) show the LDOS at $x=0, L/2, L$, and total DOS, respectively; Additional LDOS results in the bulk of the wire are presented in Fig.~\ref{fig:L_muVar5.0_3.0_2_LDOS} (Appendix~\ref{app:LDOS});
    (i)-(l) shows the topological visibility and thermal conductance from both ends. 
    }
    \label{fig:L_muVar5.0_3.0_2}
\end{figure*}

\begin{figure*}[htbp]
    \centering
    \includegraphics[width=6.8in]{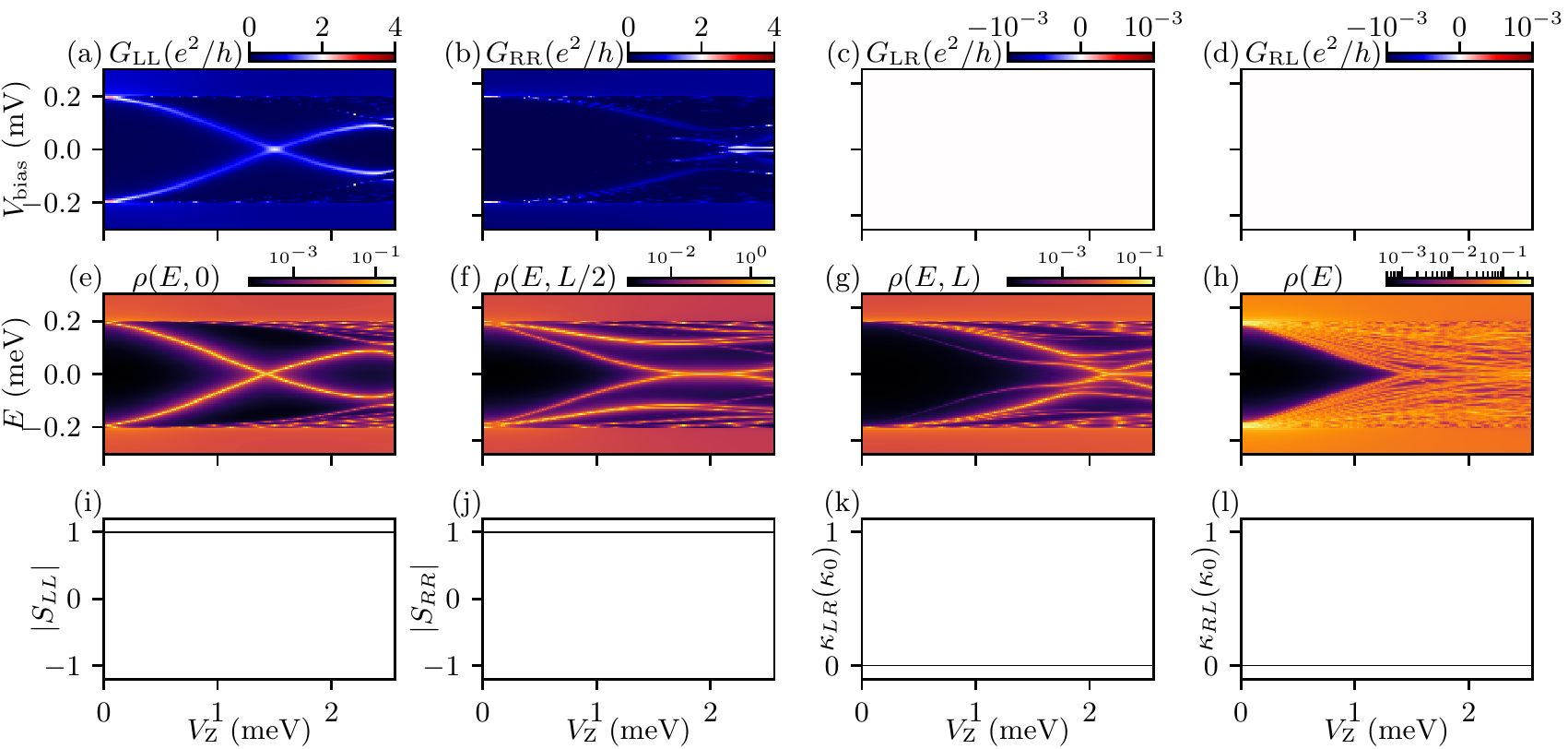}
    \caption{Ten-micron wire in the presence of intermediate disorder ($\sigma=5$ meV). 
    (a)-(d) show the local and nonlocal conductances;
    (e)-(h) show the LDOS at $x=0, L/2, L$, and total DOS, respectively; Additional LDOS results in the bulk of the wire are presented in Fig.~\ref{fig:L_muVar5.0_10.0_2_LDOS} (Appendix~\ref{app:LDOS});
    (i)-(l) shows the topological visibility and thermal conductance from both ends. 
    }
    \label{fig:L_muVar5.0_10.0_2}
\end{figure*}

On the other hand, if disorder strength continues to increase beyond $\sigma=3$ meV, we find the wire will finally enter the trivial regime.
We present two sets of results for a disorder strength of $\sigma=5$ meV in a one-micron, three-micron, and ten-micron wire as shown in Figs.~\ref{fig:L_muVar5.0_1.0}-\ref{fig:L_muVar5.0_10.0} and Figs.~\ref{fig:L_muVar5.0_1.0_2}-\ref{fig:L_muVar5.0_10.0_2}, respectively.
We find that although short wires can still manifest features akin to the situation at $\sigma=$2 - 3 meV and reflect misleading topological effects, such as the peak in the thermal conductance (see Figs.~\ref{fig:L_muVar5.0_1.0}(k)-(l) and \ref{fig:L_muVar5.0_1.0_2}(k)-(l)), the long wires essentially already enter the nontopological regime (see Figs.~\ref{fig:L_muVar5.0_10.0} and~\ref{fig:L_muVar5.0_10.0_2}). 
This transition indicates the trivial nature of the wire as disorder strength increases, in sharp contrast to the previous weaker disorder case ($\sigma=2.5$ meV), as shown in Figs.\ref{fig:L_muVar2.5_0.5}-~\ref{fig:L_muVar2.5_10.0}, which is essentially topological in the long wire limit (see Fig.~\ref{fig:L_muVar2.5_10.0}).
In fact, this is the crossover regime from intermediate to strong disorder, a regime that we will discuss in detail in the following section.

\subsection{Strong disorder}\label{sec:strong}

\begin{figure*}[htbp]
    \centering
    \includegraphics[width=6.8in]{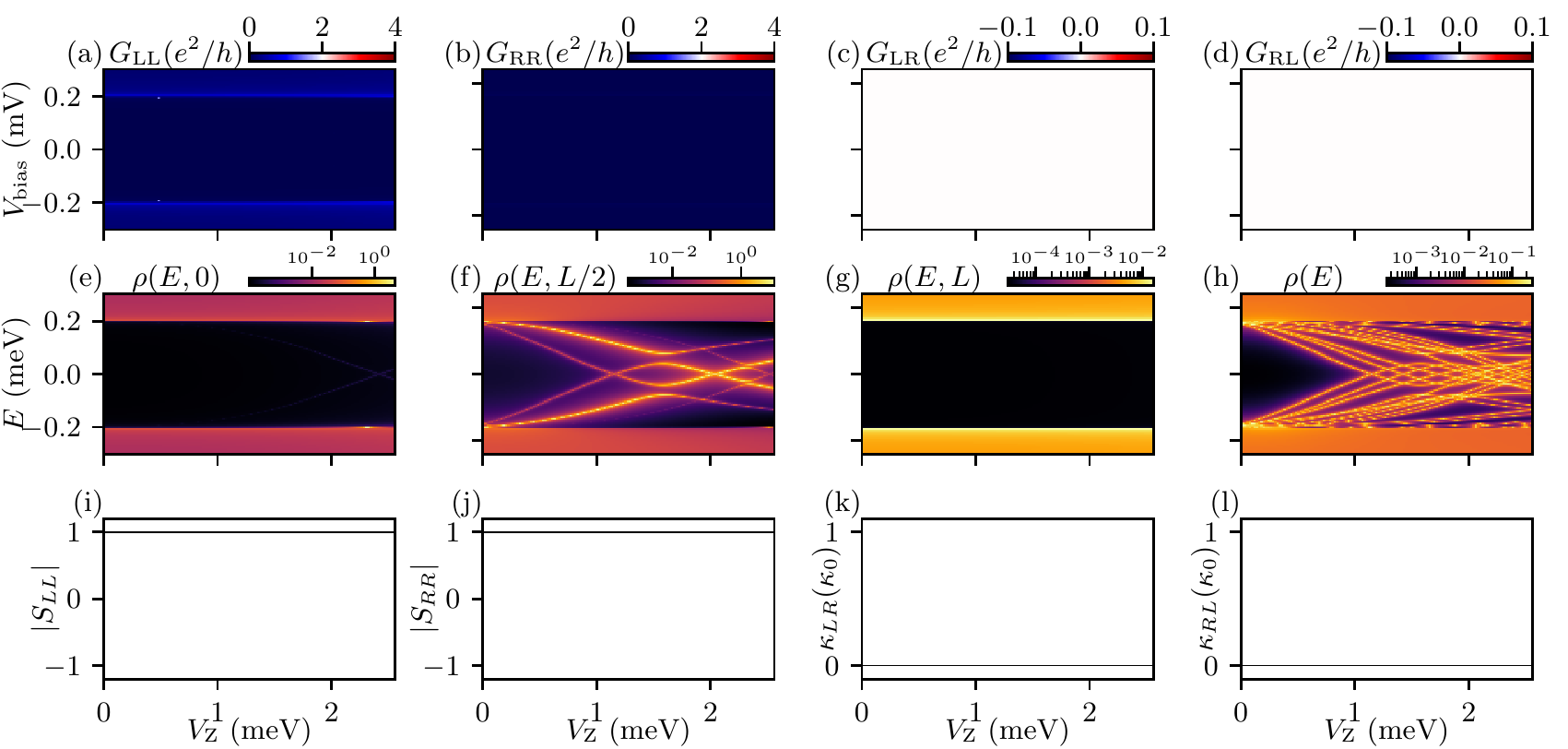}
    \caption{Three-micron wire in the presence of strong disorder ($\sigma=30$ meV) without any signals. 
    (a)-(d) show the local and nonlocal conductances;
    (e)-(h) show the LDOS at $x=0, L/2, L$, and total DOS, respectively; Additional LDOS results in the bulk of the wire are presented in Fig.~\ref{fig:L_muVar30_0_LDOS} (Appendix~\ref{app:LDOS});
    (i)-(l) shows the topological visibility and thermal conductance from both ends. 
    }
    \label{fig:L_muVar30_0}
\end{figure*}

Finally, we study the scenario of strong disorder in a relatively long wire with $L=3$ microns. In most cases with strong disorder ($\sigma=30$ meV), both local and nonlocal conductances mostly show no signals as shown in Figs.~\ref{fig:L_muVar30_0}(a)-(b) and~\ref{fig:L_muVar30_0}(c)-(d), respectively. 
Although the total DOS in Fig.~\ref{fig:L_muVar30_0}(h) indicates the existence of low-energy states due to random disorder in the bulk, these states are not necessarily localized at the wire ends, as shown in Figs.~\ref{fig:L_muVar30_0}(e)-~\ref{fig:L_muVar30_0}(g). 
Furthermore, the absence of phase transitions in the topological visibility and thermal conductances in Fig.~\ref{fig:L_muVar30_0}(i)-(l) suggests that the system remains topologically trivial in the presence of strong disorder.

\begin{figure*}[htbp]
    \centering
    \includegraphics[width=6.8in]{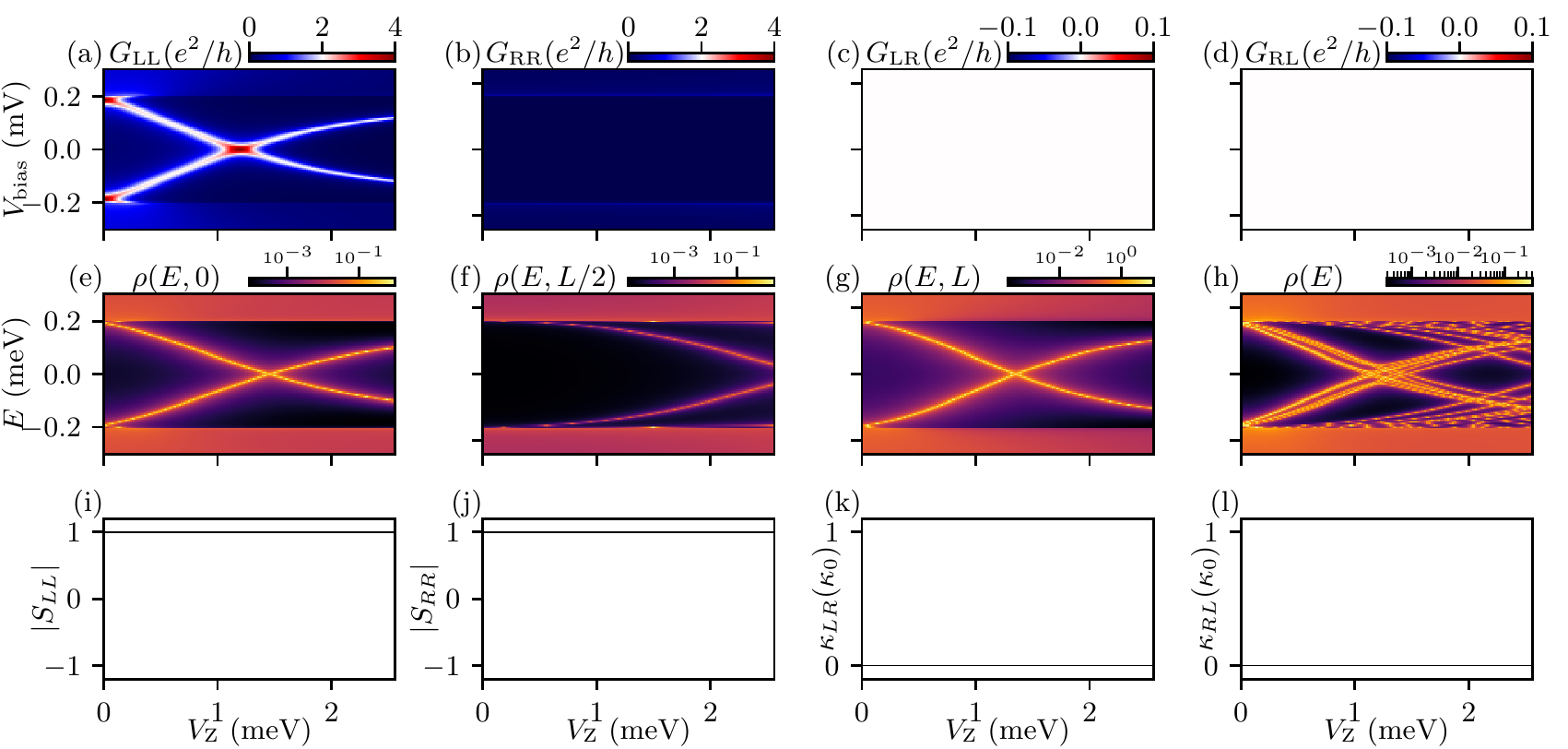}
    \caption{Three-micron wire in the presence of strong disorder ($\sigma=30$ meV) with signals in the local conductance. 
    (a)-(d) show the local and nonlocal conductances;
    (e)-(h) show the LDOS at $x=0, L/2, L$, and total DOS, respectively; Additional LDOS results in the bulk of the wire are presented in Fig.~\ref{fig:L_muVar30_1_LDOS} (Appendix~\ref{app:LDOS});
    (i)-(l) shows the topological visibility and thermal conductance from both ends. 
    }
    \label{fig:L_muVar30_1}
\end{figure*}

\begin{figure*}[htbp]
    \centering
    \includegraphics[width=6.8in]{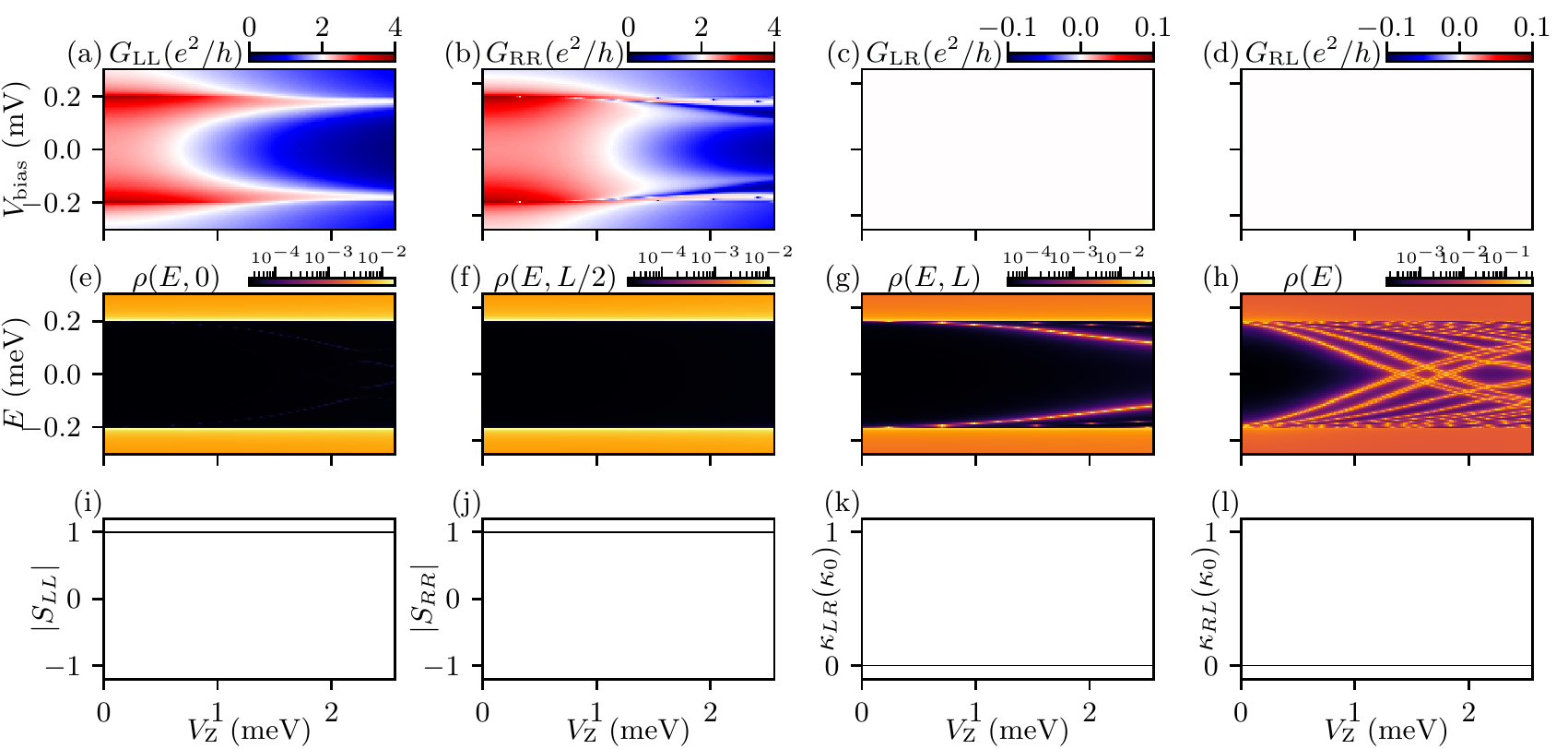}
    \caption{Three-micron wire in the presence of strong disorder ($\sigma=30$ meV) with signals in the local conductance. 
    (a)-(d) show the local and nonlocal conductances;
    (e)-(h) show the LDOS at $x=0, L/2, L$, and total DOS, respectively; Additional LDOS results in the bulk of the wire are presented in Fig.~\ref{fig:L_muVar30_2_LDOS} (Appendix~\ref{app:LDOS});
    (i)-(l) shows the topological visibility and thermal conductance from both ends. 
    }
    \label{fig:L_muVar30_2}
\end{figure*}

In some rare cases, the lowest energy state can be accidentally localized at the end of the wire, resulting in a ZBCP in the local conductance spectra, as shown in Fig.~\ref{fig:L_muVar30_1}(a)-(b). (These are what we have referred to as `ugly' ZBCPs elsewhere~\cite{pan2020physical}.)
However, this end localization does not necessarily indicate the existence of topological superconductivity. Instead, these states are just accidentally localized at the wire ends in addition to those states localized in the bulk of the wire. Through finetuning and postselection such ZBCPs may be experimentally observed, but this is not connected with topological physics and is a manifestation of disorder-induced Andreev bound states. 
The topological visibility and thermal conductance in Figs.~\ref{fig:L_muVar30_1}(i)-(l) also confirm the trivial nature of the system in the presence of strong disorder. 
In addition, we find strong disorder-induced subgap conductance features from both ends in the disorder configuration of Fig.~\ref{fig:L_muVar30_2}, which, however, are not ZBCPs. In general, Andreev bound states in the strong disorder case may induce subgap local conductance features, but this is not an indicator of topology at all.
Although finetuning and postselection can occasionally produce ZBCPs, sometimes even with approximate quantization, in the strong disorder case, this is accidental arising entirely from disorder, as we have discussed in detail elsewhere~\cite{ahn2021estimating,dassarma2021disorderinduced,pan2020physical}.

We do believe that most of the reported Microsoft data are not in this strong disorder regime, although most earlier Majorana experiments in the literature are~\cite{dassarma2023search}.

We show in Figs.~\ref{fig:L_muVar10_1} and~\ref{fig:L_muVar10_3} in Appendix~\ref{app:C} our results for the disorder strength of 10 meV, which are essentially identical to the results for the 30 meV disorder thus establishing that the strong disorder situation comes into play already for disorder $\sim$ 10 meV (or lower).  This is perhaps not surprising because this disorder strength is already roughly 50 times the induced zero field gap.

\section{Discussion and Conclusions}\label{sec:discussion}
We have provided for Majorana nanowires, as functions of disorder and system length,  detailed theoretical results for the experimentally measured local and nonlocal conductance tunnel conductance (i.e., the four components of the conductance matrix) as well as four strictly theoretical quantities; density of states, Majorana localization length, transport topological invariant, thermal conductance.  By comparing these eight different properties in the weak, intermediate, and strong disorder regimes, we conclude that the recent breakthrough Microsoft experiment is in the intermediate disorder regime where the effective disorder strength is larger than the induced superconducting gap, but perhaps not so large as to completely suppress the topological properties as in the strong disorder regime.  The weak disorder regime, characterized by a disorder strength comparable to or less than the induced gap, manifests topology essentially identical to the pristine case, reflecting the robustness of topology to local perturbations.  The intermediate disorder case is complex, with many competing factors complicating a simple interpretation of the conductance data.  There may be multiple transitions between topological and nontopological phases with increasing Zeeman splitting because of the presence of disorder-induced multiple Majorana modes throughout the bulk.  The nonlocal conductance may not manifest any gap closing and/or opening at all because of disorder. Since experimentally only conductance can be measured and not any topological invariants, it may be a challenge to infer the nature of topology in particular samples in the intermediate disorder situations although there are situations, depending sensitively on wire length and disorder configurations, where topology manifests itself clearly for intermediate disorder.

{We note that the three disorder regimes (weak/intermediate/strong) are not sharply divided, and are just smooth crossovers with their boundaries being dependent on many energy parameters of the pristine system: induced gap,  gap of the parent SC, SO coupling, chemical potential, Zeeman splitting, temperature, etc.  Many of these energy scales are unknown in the experimental nanowires, making it a challenge for determining the precise boundaries between weak/intermediate/strong disorder regimes.  This is why we provide a great deal of theoretical results varying the disorder strength so that the appropriate disorder regime can be discerned qualitatively by comparing experiment and theory directly. In general, the weak disorder regime is characterized by disorder strength being of the order of the SC gap (or less), whereas the strong disorder regime arises for disorder much larger than the induced gap.  The details, however, also depend on other energy scales in the problem, particularly, the chemical potential. This makes it impossible to give precise numbers for these different disorder regimes.  In general, when the mean free path is much larger (smaller) than the wire length, the system is weakly (strongly) disordered, but the current experiments seem to fall in the intermediate regime where the mean free path is of the order of the wire length. We also mention that in the multi-subband situation, when several subbands are occupied in the system (a situation we do not explicitly consider in the current paper), an intrinsic disorder arises because of the randomness in the energy levels associated with these subbands, further complicating the disorder estimates~\cite{pan2022random,pan2020generic}.}  

We believe that the recent Microsoft experiment mostly explores the intermediate disorder regime, whereas, by contrast, all earlier Majorana experiments were in the strong disorder regime where topology does not exist.  Future progress depends on reducing experimental disorder further so that the samples are in the weak disorder regime where topology will be manifest and obvious. Experiments going beyond conductance, where topology itself can be directly explored (e.g., thermal conductance, braiding), will be enormously helpful at this point.

\section{Acknowledgments}
The authors are grateful to Roman Lutchyn, Chetan Nayak, Jay Deep Sau, and Tudor Stanescu for many helpful discussions. This work is supported by the Laboratory for Physical Sciences.
\bibliography{Paper_TGP.bib}
\appendix
\renewcommand\thefigure{\thesection.\arabic{figure}}
\section{LDOS in the bulk of the wire}\label{app:LDOS}
\setcounter{figure}{0} 
In this appendix, we present more results for the local density of states in the bulk of the wire, located at one-third and two-thirds of the wire length, in addition to the LDOS results shown in the main text.

\begin{figure}[htbp]
    \centering
    \includegraphics[width=3.4in]{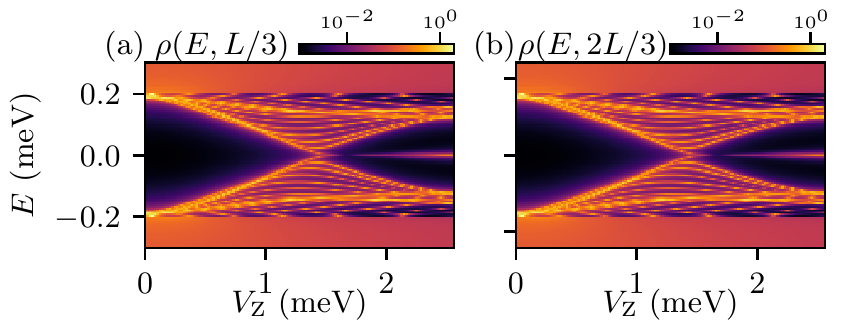}
    \caption{(a) and (b) show the LDOS at $L/3$ and $2L/3$ in a one-micron pristine wire corresponding to Fig.~\ref{fig:L_pris_1.0}.}
    \label{fig:L_pris_1.0_LDOS}
\end{figure}
\begin{figure}[htbp]
    \centering
    \includegraphics[width=3.4in]{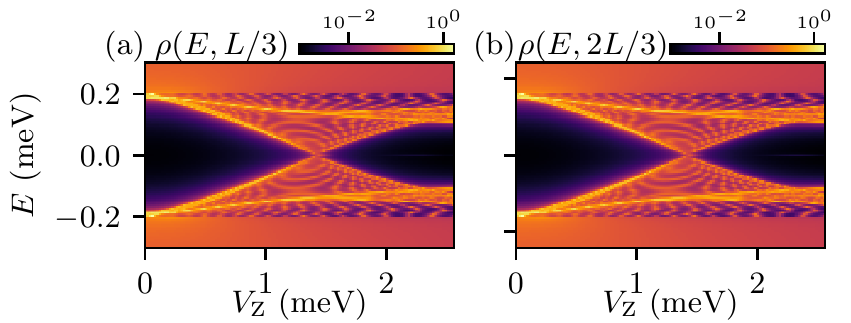}
    \caption{(a) and (b) show the LDOS at $L/3$ and $2L/3$ in a three-micron pristine wire corresponding to Fig.~\ref{fig:L_pris_3.0}.}
    \label{fig:L_pris_3.0_LDOS}
\end{figure}

\begin{figure}[htbp]
    \centering
    \includegraphics[width=3.4in]{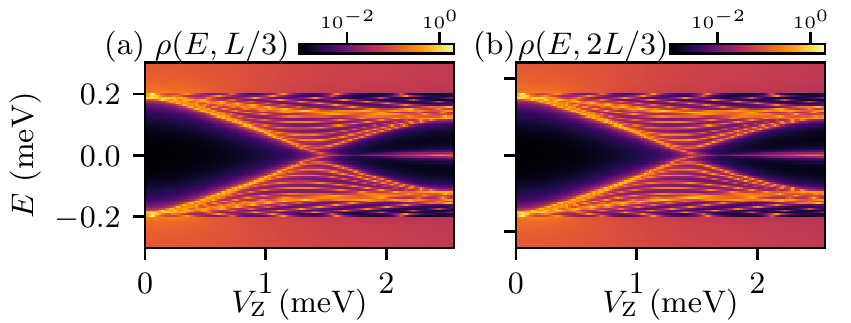}
    \caption{(a) and (b) show the LDOS at $L/3$ and $2L/3$ in a one-micron wire in the presence of weak disorder ($\sigma=0.3$ meV) corresponding to Fig.~\ref{fig:L_muVar0.3_1.0}.}
    \label{fig:L_muVar0.3_1.0_LDOS}
\end{figure}
\begin{figure}[htbp]
    \centering
    \includegraphics[width=3.4in]{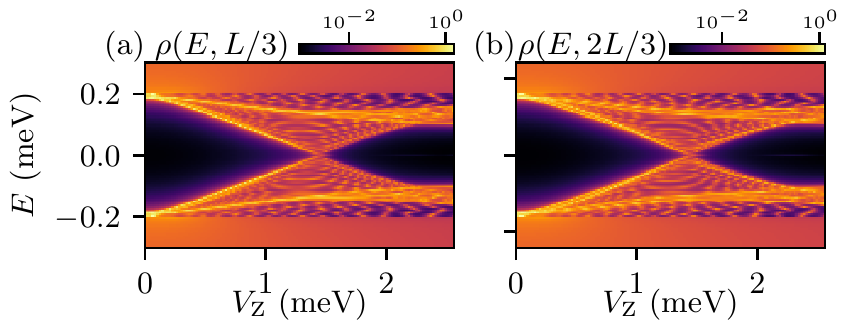}
    \caption{(a) and (b) show the LDOS at $L/3$ and $2L/3$ in a three-micron wire in the presence of weak disorder ($\sigma=0.3$ meV) corresponding to Fig.~\ref{fig:L_muVar0.3_3.0}.}
    \label{fig:L_muVar0.3_3.0_LDOS}
\end{figure}

\begin{figure}[htbp]
    \centering
    \includegraphics[width=3.4in]{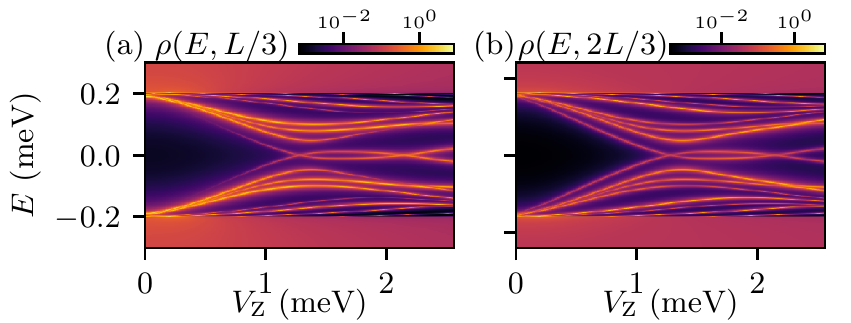}
    \caption{(a) and (b) show the LDOS at $L/3$ and $2L/3$ in a 0.5-micron wire in the presence of intermediate disorder ($\sigma=2.5$ meV) corresponding to Fig.~\ref{fig:L_muVar2.5_0.5}.}
    \label{fig:L_muVar2.5_0.5_LDOS}
\end{figure}
\begin{figure}[htbp]
    \centering
    \includegraphics[width=3.4in]{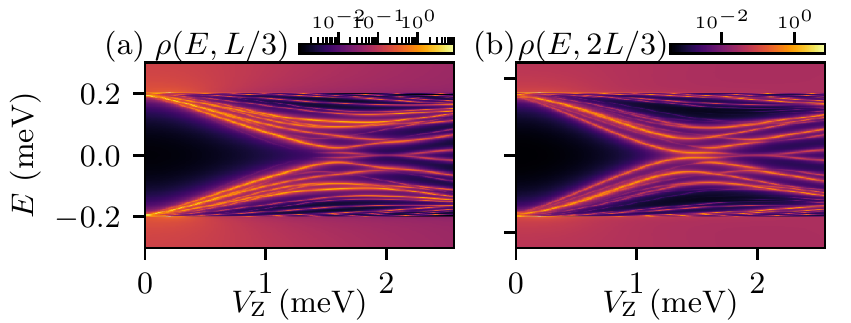}
    \caption{(a) and (b) show the LDOS at $L/3$ and $2L/3$ in a one-micron wire in the presence of intermediate disorder ($\sigma=2.5$ meV) corresponding to Fig.~\ref{fig:L_muVar2.5_1.0}.}
    \label{fig:L_muVar2.5_1.0_LDOS}
\end{figure}
\begin{figure}[htbp]
    \centering
    \includegraphics[width=3.4in]{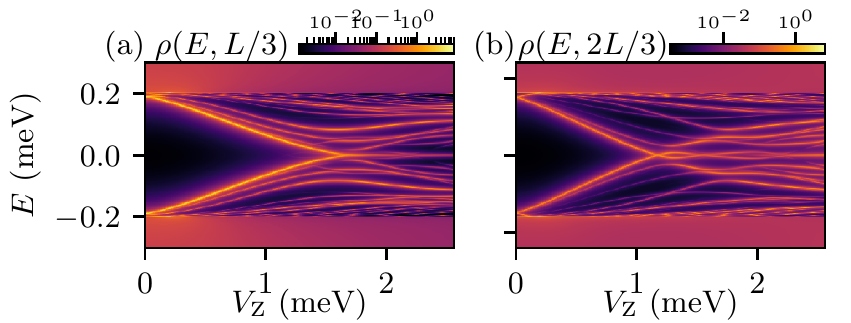}
    \caption{(a) and (b) show the LDOS at $L/3$ and $2L/3$ in a 1.5-micron wire in the presence of intermediate disorder ($\sigma=2.5$ meV) corresponding to Fig.~\ref{fig:L_muVar2.5_1.5}.}
    \label{fig:L_muVar2.5_1.5_LDOS}
\end{figure}
\begin{figure}[htbp]
    \centering
    \includegraphics[width=3.4in]{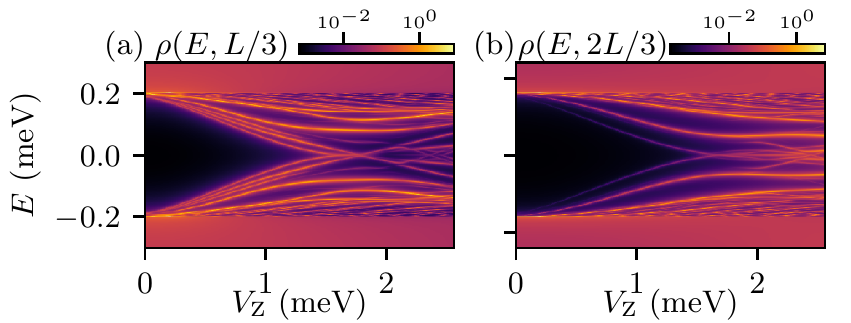}
    \caption{(a) and (b) show the LDOS at $L/3$ and $2L/3$ in a three-micron wire in the presence of intermediate disorder ($\sigma=2.5$ meV) corresponding to Fig.~\ref{fig:L_muVar2.5_3.0}.}
    \label{fig:L_muVar2.5_3.0_LDOS}
\end{figure}
\begin{figure}[htbp]
    \centering
    \includegraphics[width=3.4in]{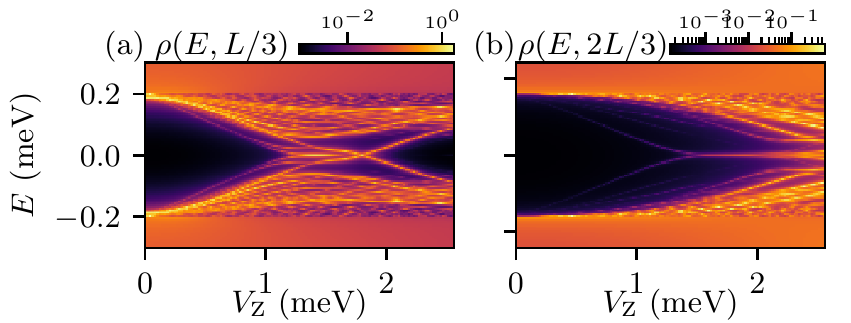}
    \caption{(a) and (b) show the LDOS at $L/3$ and $2L/3$ in a ten-micron wire in the presence of intermediate disorder ($\sigma=2.5$ meV) corresponding to Fig.~\ref{fig:L_muVar2.5_10.0}.}
    \label{fig:L_muVar2.5_10.0_LDOS}
\end{figure}
\begin{figure}[htbp]
    \centering
    \includegraphics[width=3.4in]{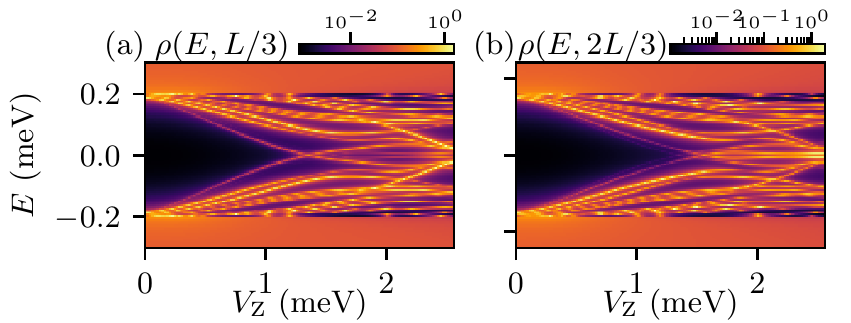}
    \caption{(a) and (b) show the LDOS at $L/3$ and $2L/3$ in a one-micron wire in the presence of intermediate disorder ($\sigma=3$ meV) corresponding to Fig.~\ref{fig:L_muVar3.0_1.0}.}
    \label{fig:L_muVar3.0_1.0_LDOS}
\end{figure}
\begin{figure}[htbp]
    \centering
    \includegraphics[width=3.4in]{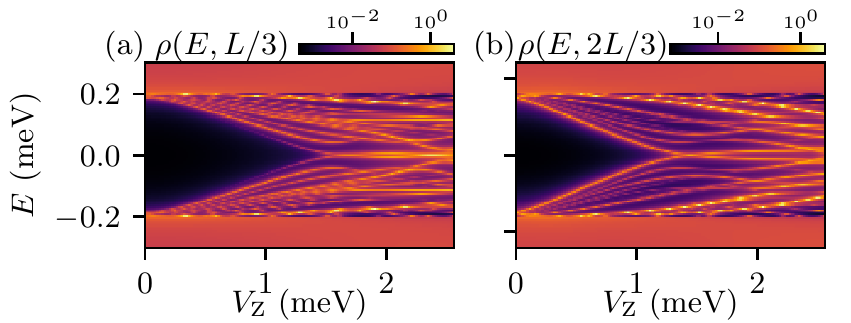}
    \caption{(a) and (b) show the LDOS at $L/3$ and $2L/3$ in a two-micron wire in the presence of intermediate disorder ($\sigma=3$ meV) corresponding to Fig.~\ref{fig:L_muVar3.0_2.0}.}
    \label{fig:L_muVar3.0_2.0_LDOS}
\end{figure}
\begin{figure}[htbp]
    \centering
    \includegraphics[width=3.4in]{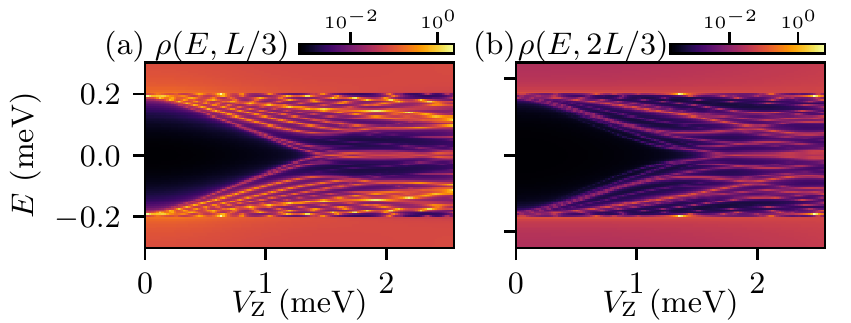}
    \caption{(a) and (b) show the LDOS at $L/3$ and $2L/3$ in a three-micron wire in the presence of intermediate disorder ($\sigma=3$ meV) corresponding to Fig.~\ref{fig:L_muVar3.0_3.0}.}
    \label{fig:L_muVar3.0_3.0_LDOS}
\end{figure}
\begin{figure}[htbp]
    \centering
    \includegraphics[width=3.4in]{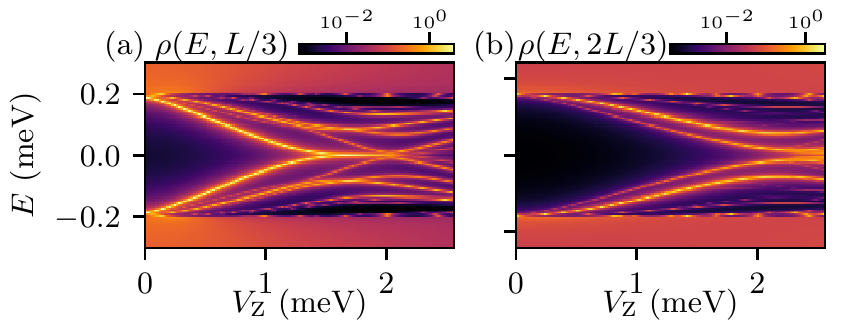}
    \caption{(a) and (b) show the LDOS at $L/3$ and $2L/3$ in a one-micron wire in the presence of intermediate disorder ($\sigma=5$ meV) corresponding to Fig.~\ref{fig:L_muVar5.0_1.0}.}
    \label{fig:L_muVar5.0_1.0_LDOS}
\end{figure}
\begin{figure}[htbp]
    \centering
    \includegraphics[width=3.4in]{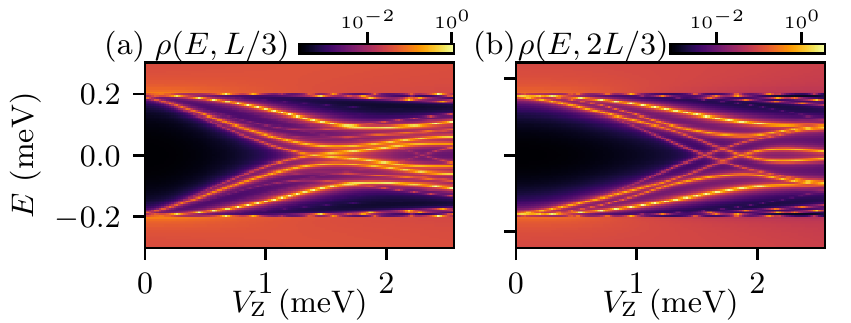}
    \caption{(a) and (b) show the LDOS at $L/3$ and $2L/3$ in a three-micron wire in the presence of intermediate disorder ($\sigma=5$ meV) corresponding to Fig.~\ref{fig:L_muVar5.0_3.0}.}
    \label{fig:L_muVar5.0_3.0_LDOS}
\end{figure}
\begin{figure}[htbp]
    \centering
    \includegraphics[width=3.4in]{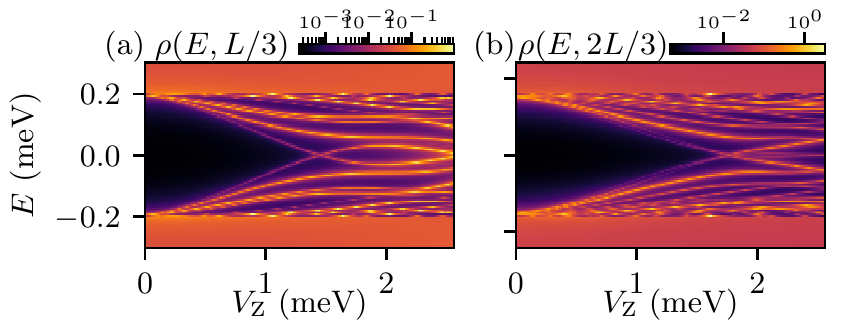}
    \caption{(a) and (b) show the LDOS at $L/3$ and $2L/3$ in a ten-micron wire in the presence of intermediate disorder ($\sigma=5$ meV) corresponding to Fig.~\ref{fig:L_muVar5.0_10.0}.}
    \label{fig:L_muVar5.0_10.0_LDOS}
\end{figure}
    
\begin{figure}[htbp]
    \centering
    \includegraphics[width=3.4in]{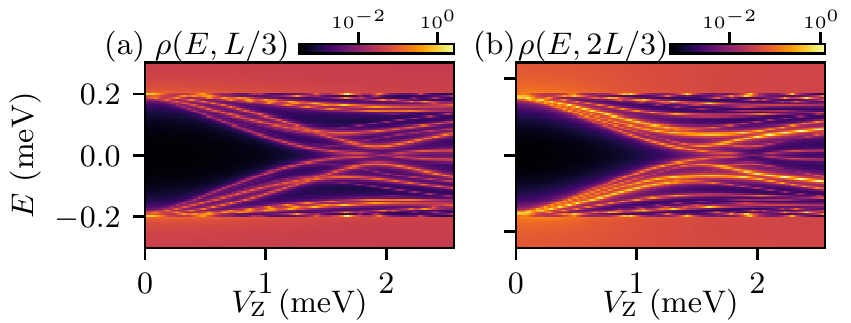}
    \caption{(a) and (b) show the LDOS at $L/3$ and $2L/3$ in a one-micron wire in the presence of intermediate disorder ($\sigma=5$ meV) corresponding to Fig.~\ref{fig:L_muVar5.0_1.0_2}.}
    \label{fig:L_muVar5.0_1.0_2_LDOS}
\end{figure}
\begin{figure}[htbp]
    \centering
    \includegraphics[width=3.4in]{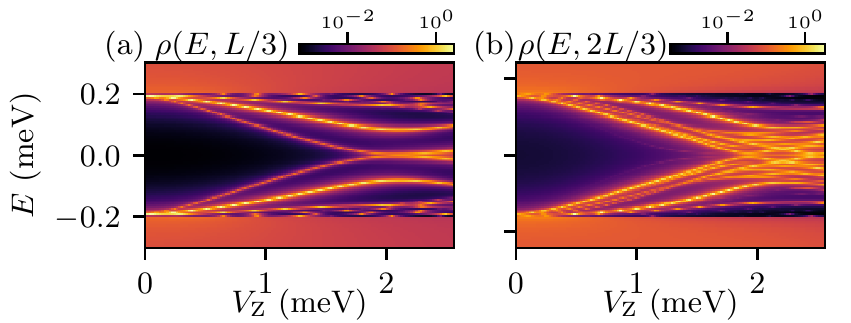}
    \caption{(a) and (b) show the LDOS at $L/3$ and $2L/3$ in a three-micron wire in the presence of intermediate disorder ($\sigma=5$ meV) corresponding to Fig.~\ref{fig:L_muVar5.0_3.0_2}.}
    \label{fig:L_muVar5.0_3.0_2_LDOS}
\end{figure}
\begin{figure}[htbp]
    \centering
    \includegraphics[width=3.4in]{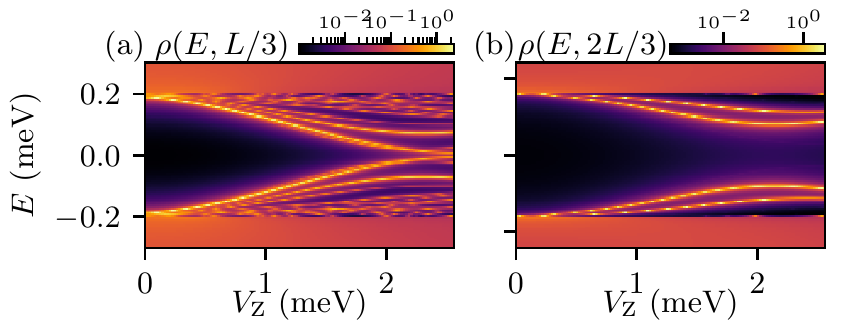}
    \caption{(a) and (b) show the LDOS at $L/3$ and $2L/3$ in a ten-micron wire in the presence of intermediate disorder ($\sigma=5$ meV) corresponding to Fig.~\ref{fig:L_muVar5.0_10.0_2}.}
    \label{fig:L_muVar5.0_10.0_2_LDOS}
\end{figure}

\begin{figure}[htbp]
    \centering
    \includegraphics[width=3.4in]{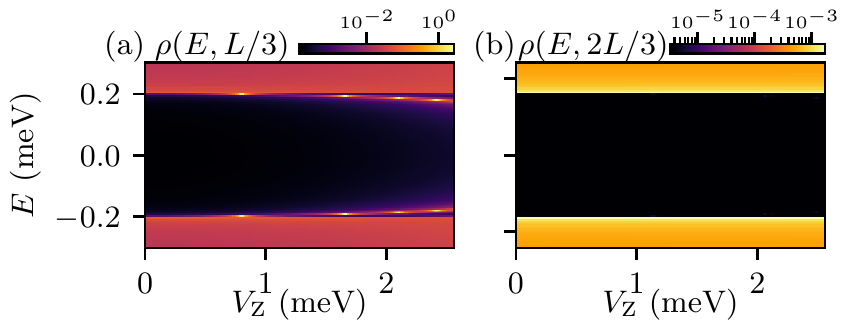}
    \caption{(a) and (b) show the LDOS at $L/3$ and $2L/3$ in a three-micron wire in the presence of strong disorder ($\sigma=30$ meV) corresponding to Fig.~\ref{fig:L_muVar30_0}.}
    \label{fig:L_muVar30_0_LDOS}
\end{figure}
\begin{figure}[htbp]
    \centering
    \includegraphics[width=3.4in]{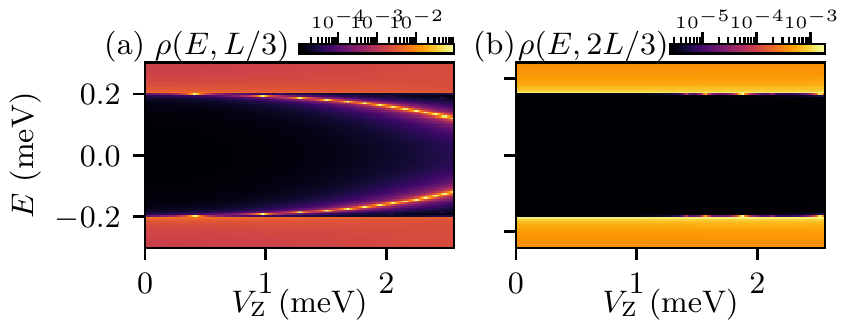}
    \caption{(a) and (b) show the LDOS at $L/3$ and $2L/3$ in a three-micron wire in the presence of strong disorder ($\sigma=30$ meV) corresponding to Fig.~\ref{fig:L_muVar30_1}.}
    \label{fig:L_muVar30_1_LDOS}
\end{figure}
\begin{figure}[htbp]
    \centering
    \includegraphics[width=3.4in]{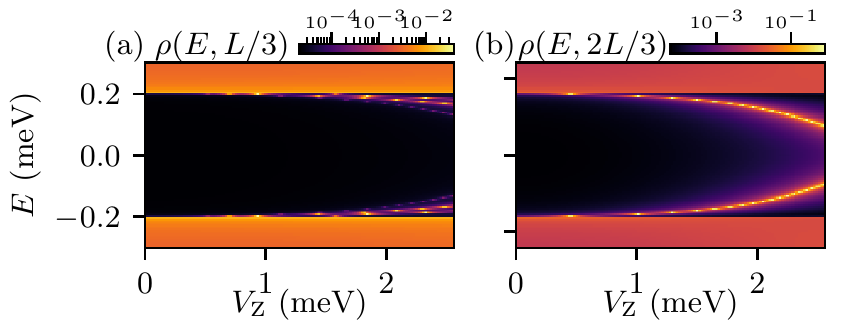}
    \caption{(a) and (b) show the LDOS at $L/3$ and $2L/3$ in a three-micron wire in the presence of strong disorder ($\sigma=30$ meV) corresponding to Fig.~\ref{fig:L_muVar30_2}.}
    \label{fig:L_muVar30_2_LDOS}
\end{figure}

\section{Wavefunctions and localization length}\label{app:localization}
\setcounter{figure}{0} 
In this appendix, we present the corresponding wavefunctions and their localization lengths. In  Figs.~\ref{fig:wf_pris_1.0}-~\ref{fig:wf_muVar3.0_3.0}, we show the amplitude of the wavefunction of the lowest state as $\abs{\Psi}^2$ (solid black line). Additionally, we decompose $\Psi$ into the two Majorana basis, with ${\gamma_1}$ (red solid line) and ${\gamma_2}$ (blue solid line) representing each basis. The localization length is estimated by fitting the spatial profile of the envelope of $\abs{\gamma_1}^2$ (red dashed line) and $\abs{\gamma_2}^2$ (blue dashed line) using an exponential function as indicated by the dashed line. For some figures (e.g., Fig.~\ref{fig:wf_muVar0.3_1.0}(b)), the absence of the dashed line is because the state is not localized at the wire ends.


\begin{figure*}[htbp]
    \centering
    \includegraphics[width=6.8in]{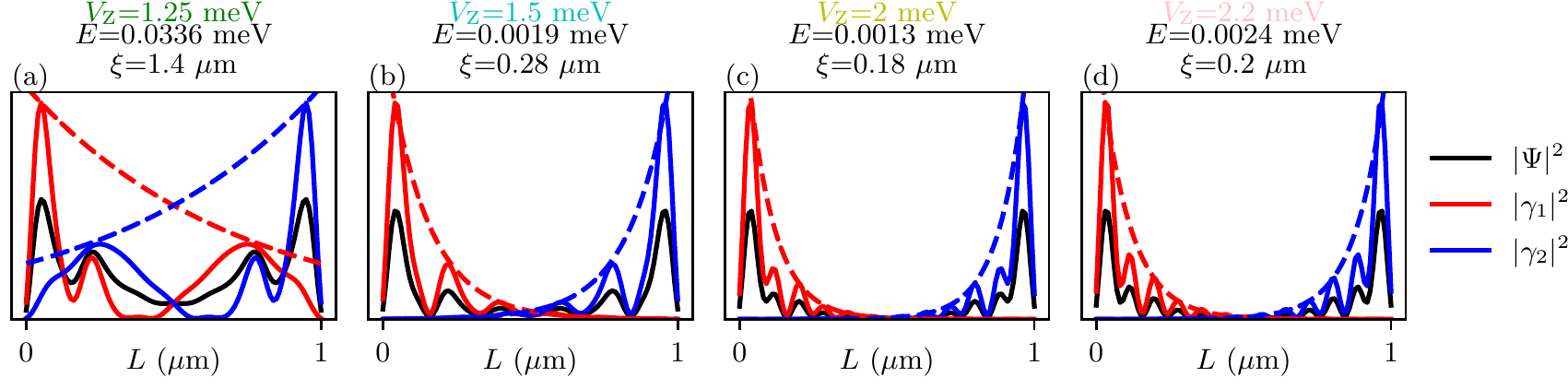}
    \caption{
    The Majorana localization lengths are shown for four different $V_\text{Z}$ (1.25 meV, 1.5 meV, 2 meV, and 2.2 meV), corresponding to four line-cuts (green, cyan, yellow, and pink) in a one-micron pristine wire (see Fig.~\ref{fig:L_pris_1.0}). The energy of the lowest subgap state is denoted by $E$.}
    \label{fig:wf_pris_1.0}
\end{figure*}

\begin{figure*}[htbp]
    \centering
    \includegraphics[width=6.8in]{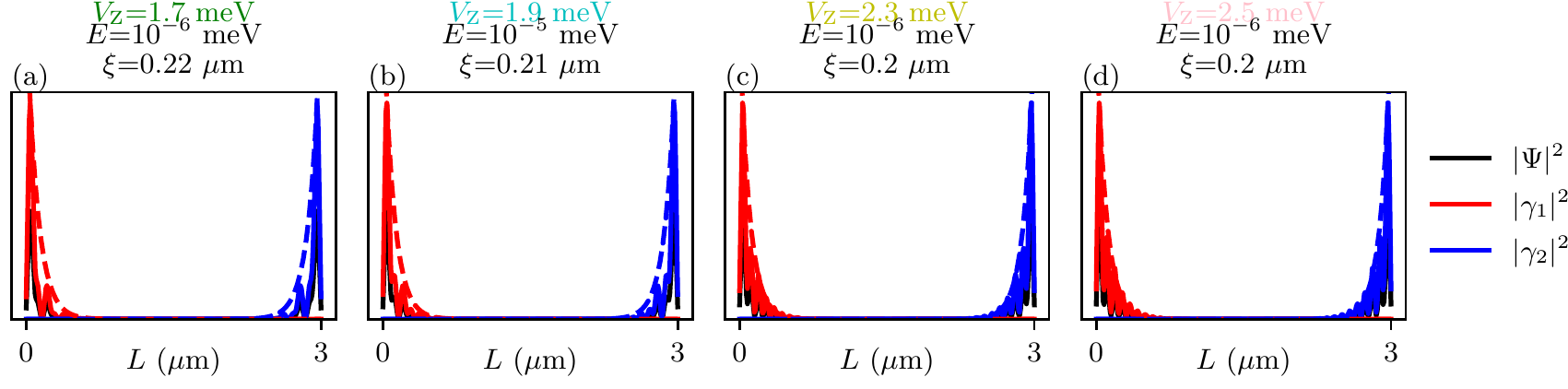}
    \caption{
    The Majorana localization lengths are shown for four different $V_\text{Z}$ (1.7 meV, 1.9 meV, 2.3 meV, and 2.5 meV), corresponding to four line-cuts (green, cyan, yellow, and pink) in a three-micron pristine wire (see Fig.~\ref{fig:L_pris_3.0}). The energy of the lowest subgap state is denoted by $E$.}
    \label{fig:wf_pris_3.0}
\end{figure*}
\begin{figure*}[htbp]
    \centering
    \includegraphics[width=6.8in]{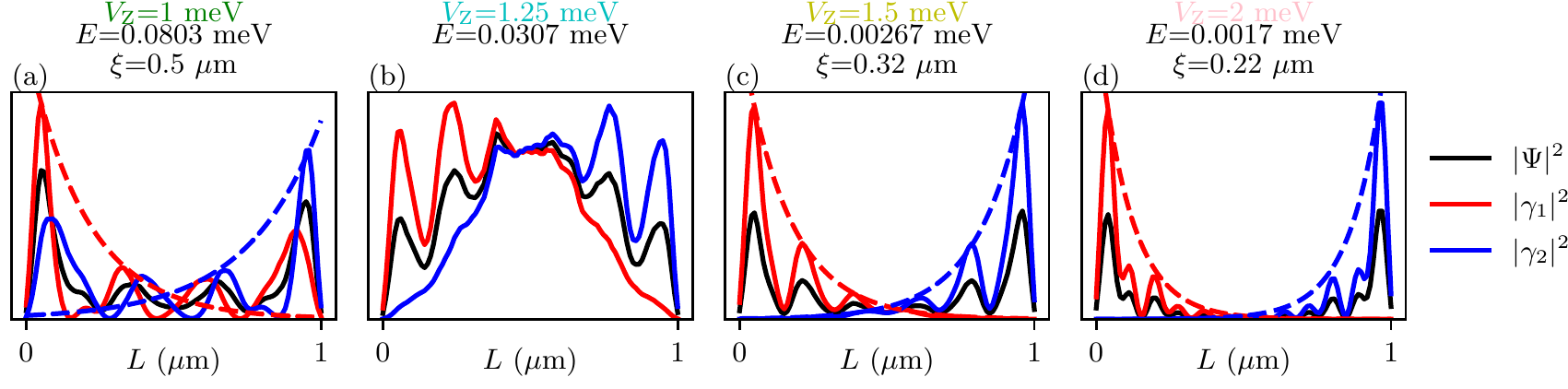}
    \caption{
    The Majorana localization lengths are shown for four different $V_\text{Z}$ (1 meV, 1.25 meV, 1.5 meV, and 2 meV), corresponding to four line-cuts (green, cyan, yellow, and pink) in a one-micron wire in the presence of weak disorder ($\sigma=0.3$ meV, see Fig.~\ref{fig:L_muVar0.3_1.0}). The energy of the lowest subgap state is denoted by $E$.}
    \label{fig:wf_muVar0.3_1.0}
\end{figure*}
\begin{figure*}[htbp]
    \centering
    \includegraphics[width=6.8in]{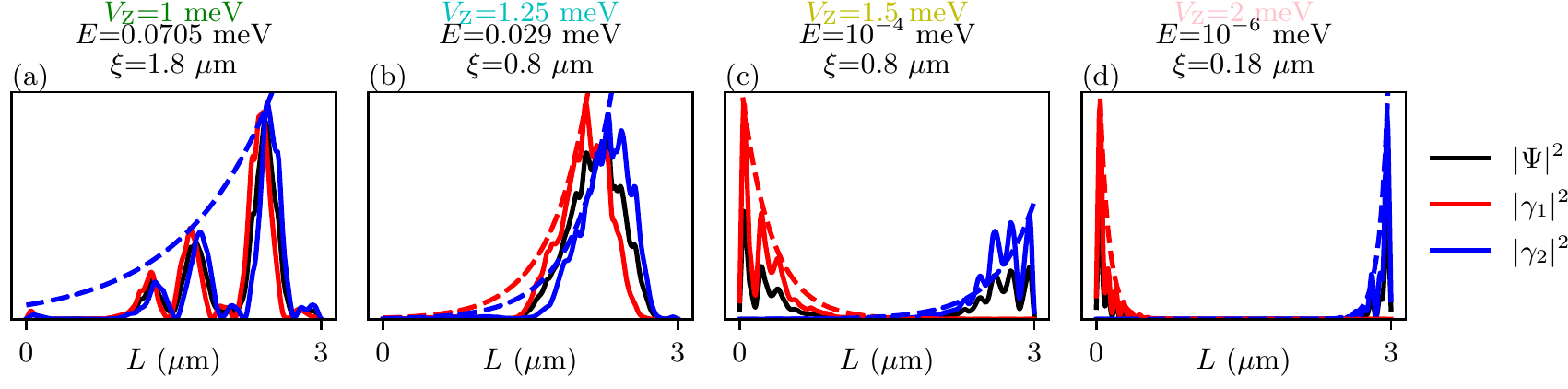}
    \caption{
    The Majorana localization lengths are shown for four different $V_\text{Z}$ (1 meV, 1.25 meV, 1.5 meV, and 2 meV), corresponding to four line-cuts (green, cyan, yellow, and pink) in a three-micron wire in the presence of weak disorder ($\sigma=0.3$ meV, see Fig.~\ref{fig:L_muVar0.3_3.0}). The energy of the lowest subgap state is denoted by $E$.}
    \label{fig:wf_muVar0.3_3.0}
\end{figure*}
\begin{figure*}[htbp]
    \centering
    \includegraphics[width=6.8in]{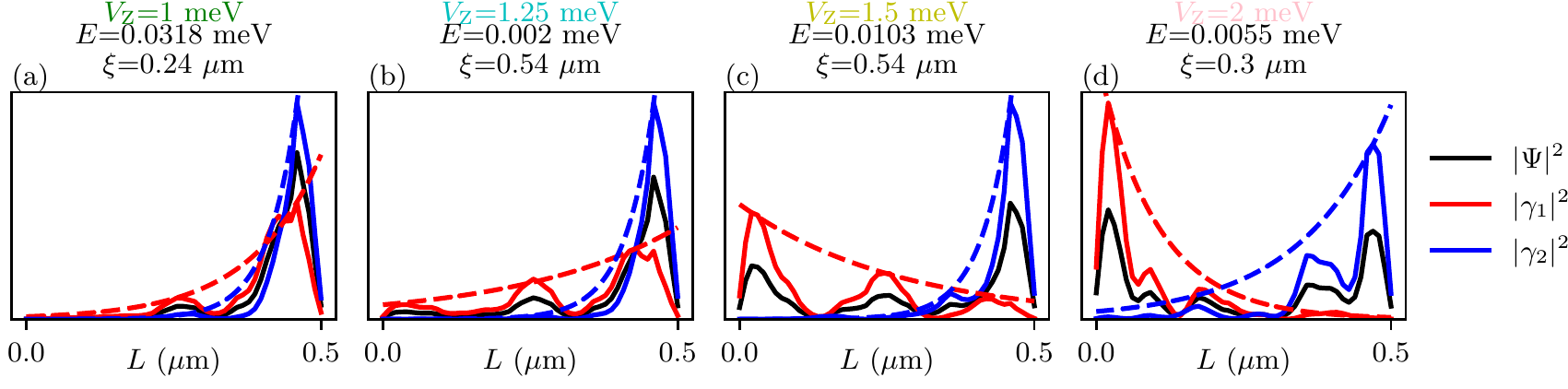}
    \caption{
    The Majorana localization lengths are shown for four different $V_\text{Z}$ (1 meV, 1.25 meV, 1.5 meV, and 2 meV), corresponding to four line-cuts (green, cyan, yellow, and pink) in a 0.5-micron wire in the presence of intermediate disorder ($\sigma=2.5$ meV, see Fig.~\ref{fig:L_muVar2.5_0.5}). The energy of the lowest subgap state is denoted by $E$.}
    \label{fig:wf_muVar2.5_0.5}
\end{figure*}
\begin{figure*}[htbp]
    \centering
    \includegraphics[width=6.8in]{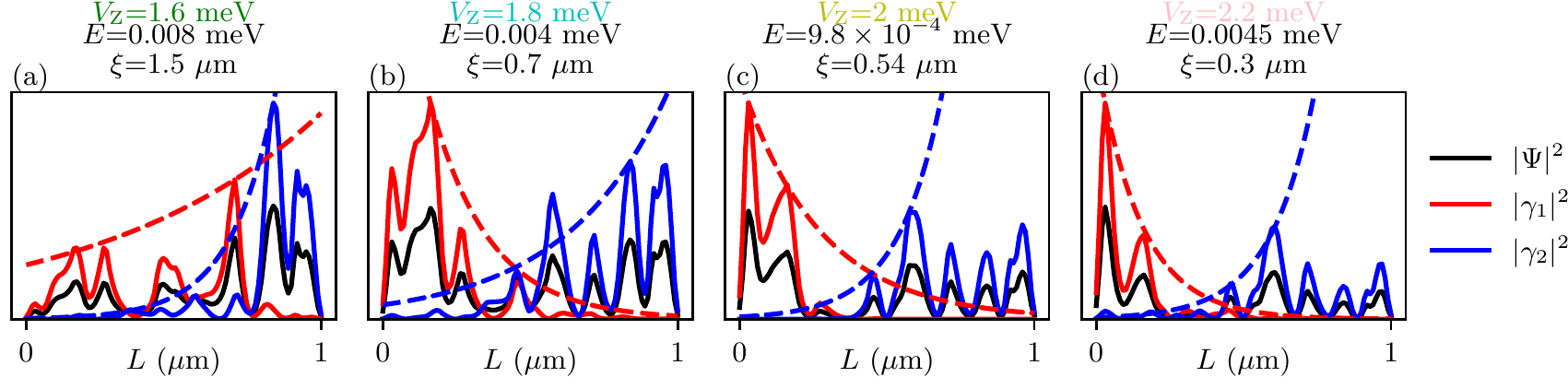}
    \caption{
    The Majorana localization lengths are shown for four different $V_\text{Z}$ (1.6 meV, 1.8 meV, 2 meV, and 2.2 meV), corresponding to four line-cuts (green, cyan, yellow, and pink) in a one-micron wire in the presence of intermediate disorder ($\sigma=2.5$ meV, see Fig.~\ref{fig:L_muVar2.5_1.0}). The energy of the lowest subgap state is denoted by $E$.}
    \label{fig:wf_muVar2.5_1.0}
\end{figure*}
\begin{figure*}[htbp]
    \centering
    \includegraphics[width=6.8in]{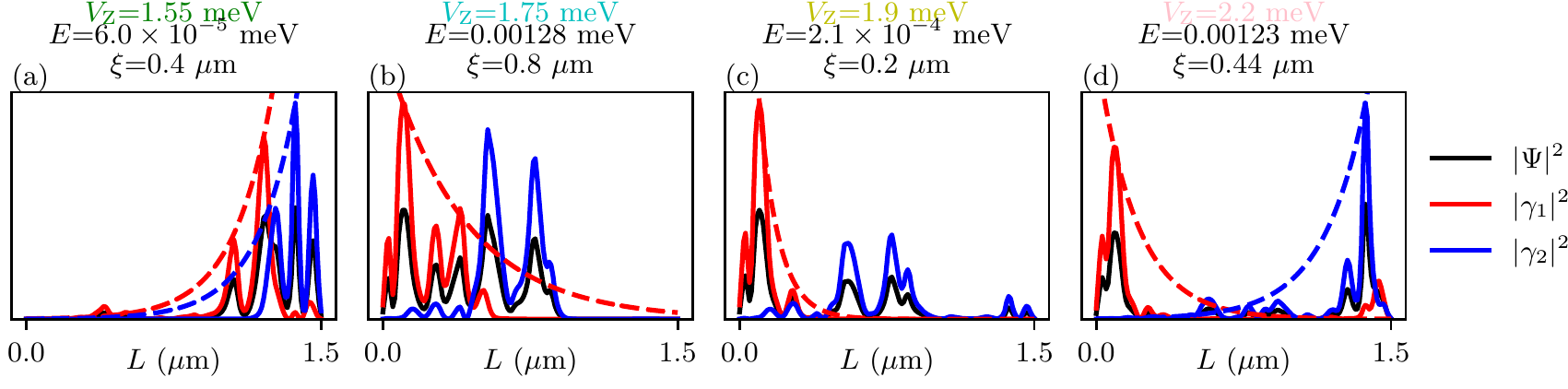}
    \caption{
    The Majorana localization lengths are shown for four different $V_\text{Z}$ (1.55 meV, 1.75 meV, 1.9 meV, and 2.2 meV), corresponding to four line-cuts (green, cyan, yellow, and pink) in a 1.5-micron wire in the presence of intermediate disorder ($\sigma=2.5$ meV, see Fig.~\ref{fig:L_muVar2.5_1.5}). The energy of the lowest subgap state is denoted by $E$.}
    \label{fig:wf_muVar2.5_1.5}
\end{figure*}

\begin{figure*}[htbp]
    \centering
    \includegraphics[width=6.8in]{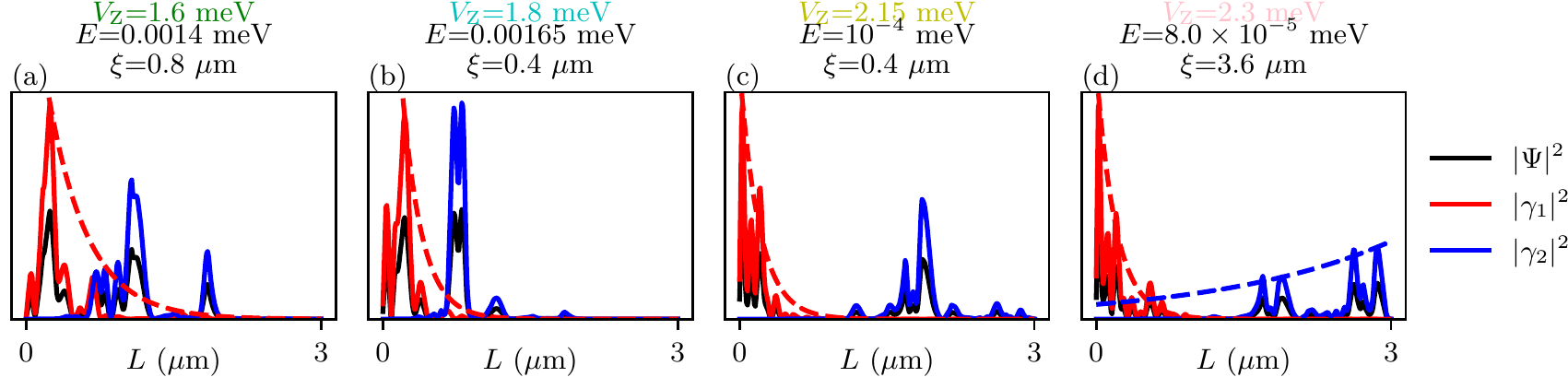}
    \caption{
    The Majorana localization lengths are shown for four different $V_\text{Z}$ (1.6 meV, 1.8 meV, 2.15 meV, and 2.3 meV), corresponding to four line-cuts (green, cyan, yellow, and pink) in a three-micron wire in the presence of intermediate disorder ($\sigma=2.5$ meV, see Fig.~\ref{fig:L_muVar2.5_3.0}). The energy of the lowest subgap state is denoted by $E$.}
    \label{fig:wf_muVar2.5_3.0}
\end{figure*}

\begin{figure*}[htbp]
    \centering
    \includegraphics[width=6.8in]{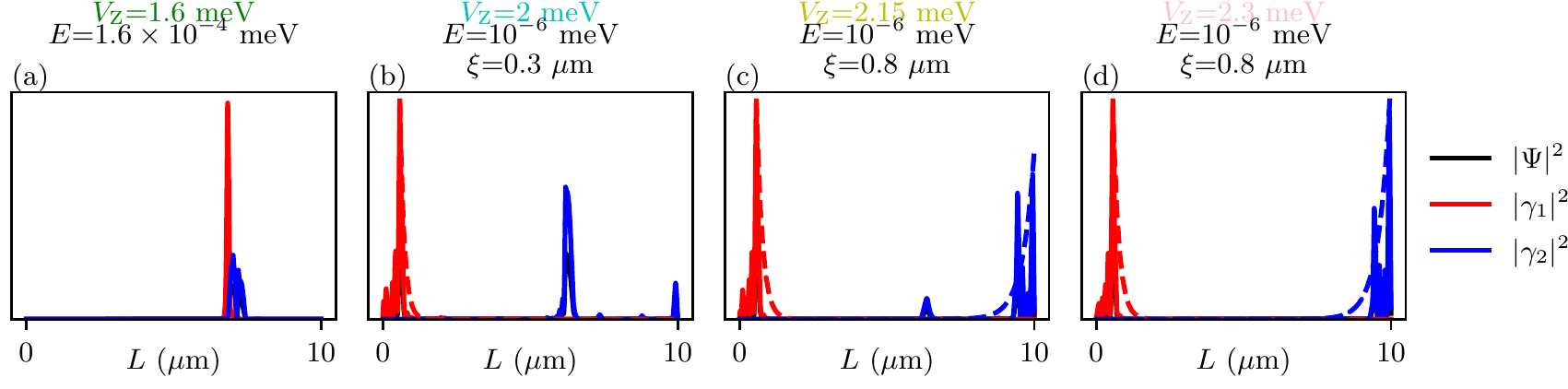}
    \caption{
    The Majorana localization lengths are shown for four different $V_\text{Z}$ (1.6 meV, 2 meV, 2.15 meV, and 2.3 meV), corresponding to four line-cuts (green, cyan, yellow, and pink) in a ten-micron wire in the presence of intermediate disorder ($\sigma=2.5$ meV, see Fig.~\ref{fig:L_muVar2.5_10.0}). The energy of the lowest subgap state is denoted by $E$.}
    \label{fig:wf_muVar2.5_10.0}
\end{figure*}

\begin{figure*}[htbp]
    \centering
    \includegraphics[width=6.8in]{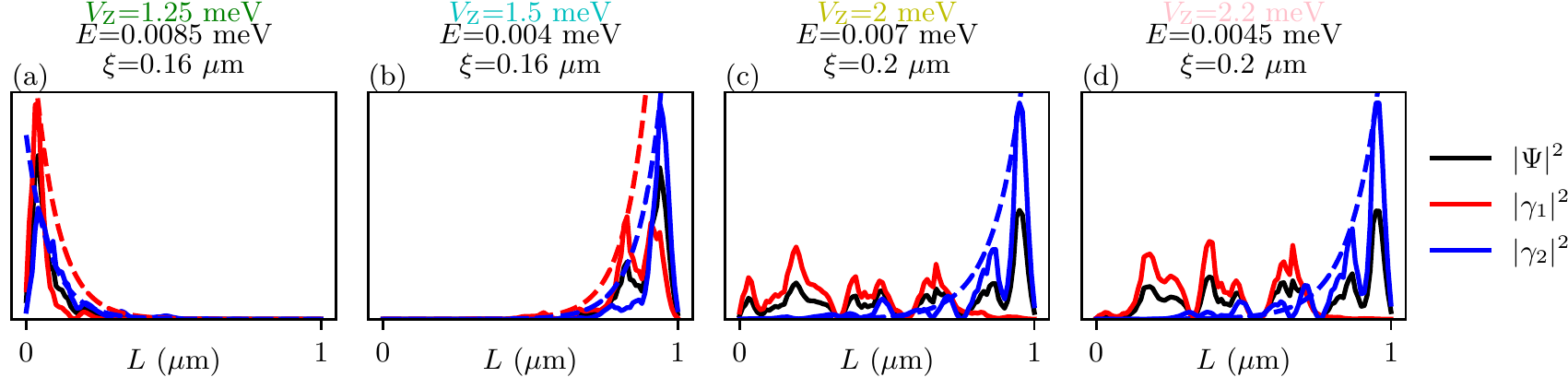}
    \caption{
    The Majorana localization lengths are shown for four different $V_\text{Z}$ (1.25 meV, 1.5 meV, 2 meV, and 2.2 meV), corresponding to four line-cuts (green, cyan, yellow, and pink) in a one-micron wire in the presence of intermediate disorder ($\sigma=3$ meV, see Fig.~\ref{fig:L_muVar3.0_1.0}). The energy of the lowest subgap state is denoted by $E$.}
    \label{fig:wf_muVar3.0_1.0}
\end{figure*}
\begin{figure*}[htbp]
    \centering
    \includegraphics[width=6.8in]{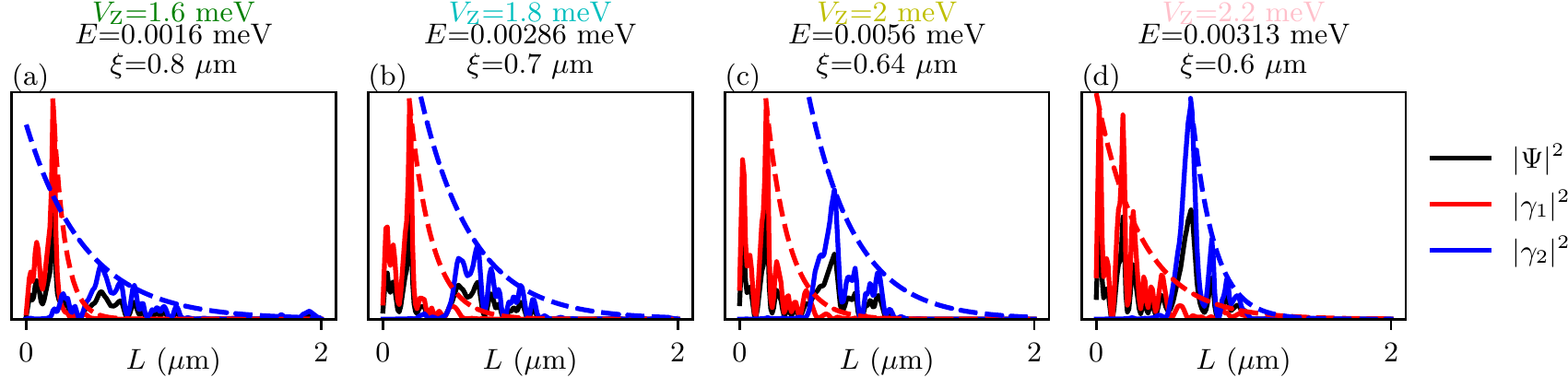}
    \caption{
    The Majorana localization lengths are shown for four different $V_\text{Z}$ (1.6 meV, 1.8 meV, 2 meV, and 2.2 meV), corresponding to four line-cuts (green, cyan, yellow, and pink) in a two-micron wire in the presence of intermediate disorder ($\sigma=3$ meV, see Fig.~\ref{fig:L_muVar3.0_2.0}). The energy of the lowest subgap state is denoted by $E$.}
    \label{fig:wf_muVar3.0_2.0}
\end{figure*}
\begin{figure*}[htbp]
    \centering
    \includegraphics[width=6.8in]{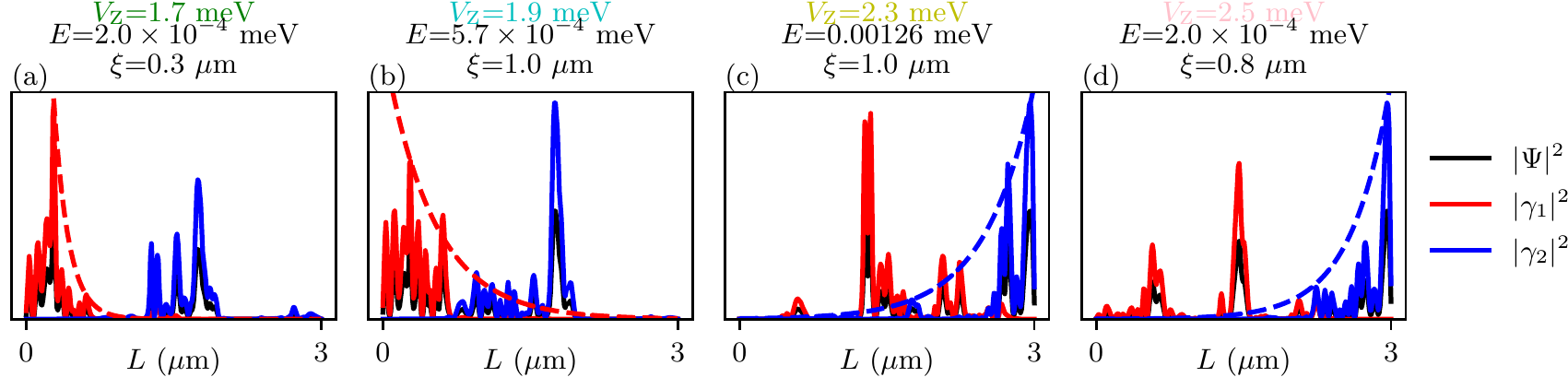}
    \caption{
    The Majorana localization lengths are shown for four different $V_\text{Z}$ (1.7 meV, 1.9 meV, 2.3 meV, and 2.5 meV), corresponding to four line-cuts (green, cyan, yellow, and pink) in a three-micron wire in the presence of intermediate disorder ($\sigma=3$ meV, see Fig.~\ref{fig:L_muVar3.0_3.0}). The energy of the lowest subgap state is denoted by $E$.}
    \label{fig:wf_muVar3.0_3.0}
\end{figure*}

\section{Additional results in the presence of strong disorder}\label{app:C}
\setcounter{figure}{0} 
In this appendix, we additionally show two sets of results 
in the presence of a fixed disorder configuration with $\sigma=10$ meV for a one-micron short wire and a three-micron long wire. These results are consistent with those discussed in the main text but use a smaller disorder strength.
\begin{figure*}[htbp]
    \centering
    \includegraphics[width=6.8in]{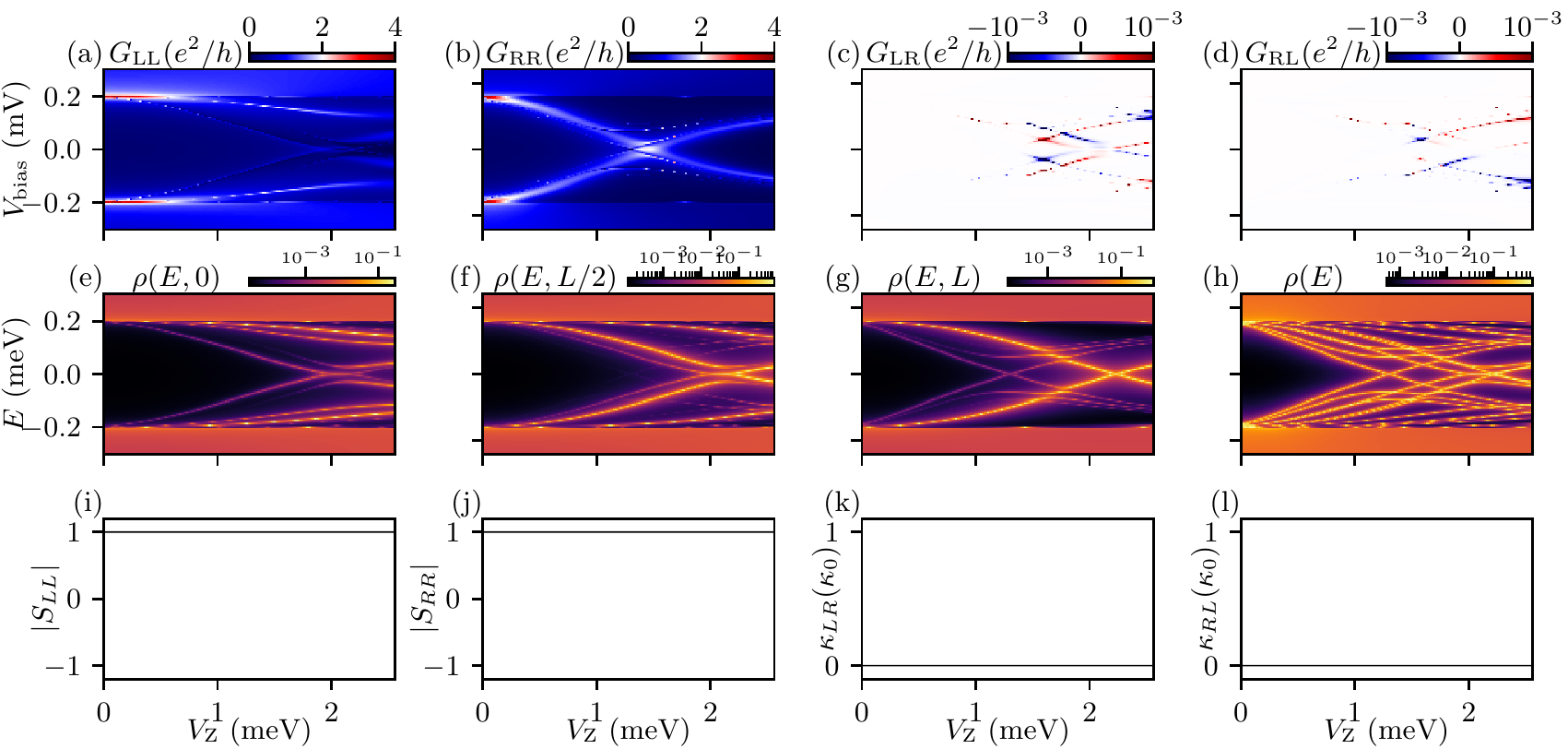}
    \caption{One-micron wire in the presence of strong disorder ($\sigma=10$ meV)
    (a)-(d) show the local and nonlocal conductances;
    (e)-(h) show the LDOS at $x=0, L/2, L$, and total DOS, respectively;
    (i)-(l) shows the topological visibility and thermal conductance from both ends. 
    }
    \label{fig:L_muVar10_1}
\end{figure*}
\begin{figure*}[htbp]
    \centering
    \includegraphics[width=6.8in]{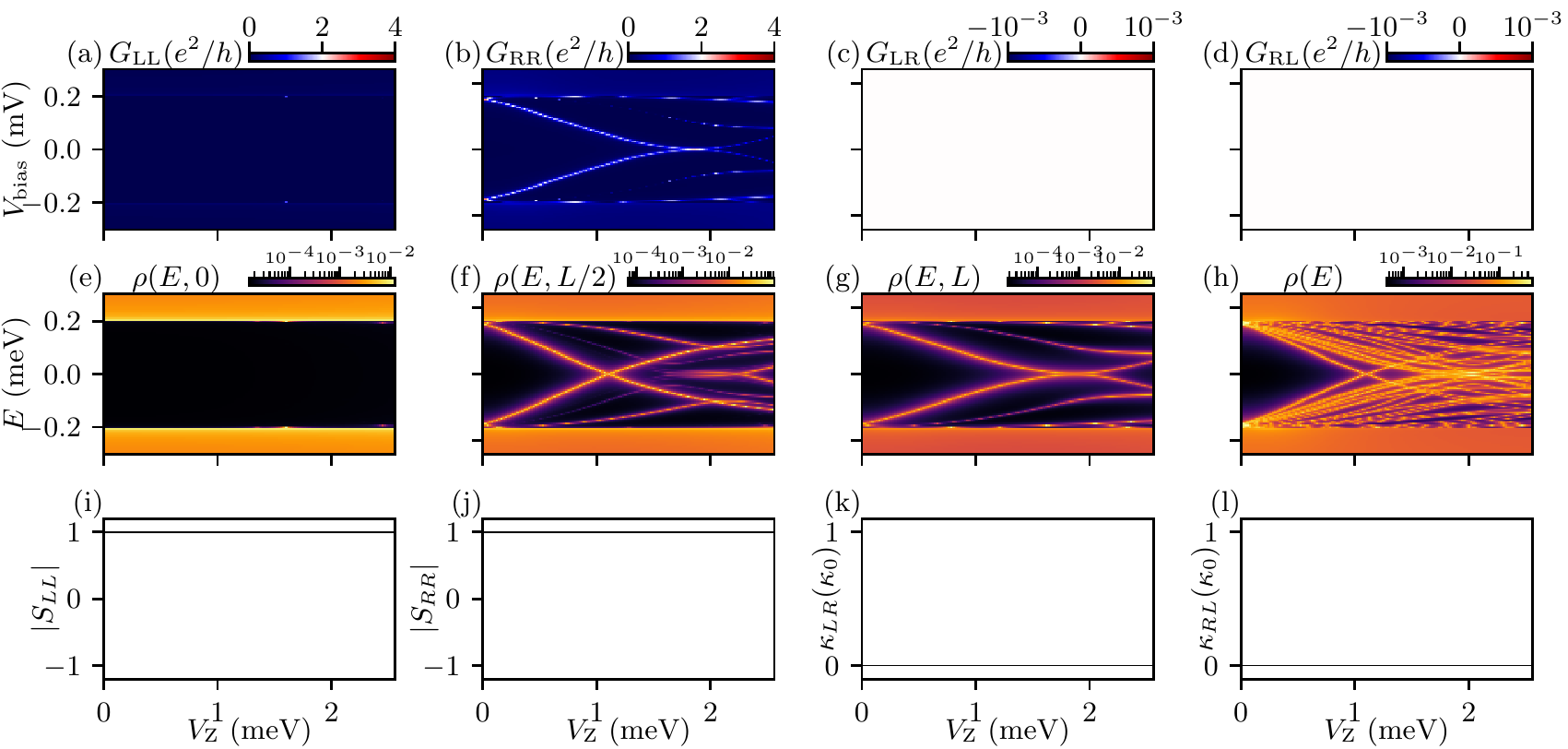}
    \caption{Three-micron wire in the presence of strong disorder ($\sigma=10$ meV)
    (a)-(d) show the local and nonlocal conductances;
    (e)-(h) show the LDOS at $x=0, L/2, L$, and total DOS, respectively;
    (i)-(l) shows the topological visibility and thermal conductance from both ends. 
    }
    \label{fig:L_muVar10_3}
\end{figure*}

\end{document}